\documentclass{emulateapj}

\def\etal{{et~al. \,}}

\def\sles{\lower2pt\hbox{$\buildrel {\scriptstyle <}
   \over {\scriptstyle\sim}$}}

\def\sgreat{\lower2pt\hbox{$\buildrel {\scriptstyle >}
   \over {\scriptstyle\sim}$}}
\def\sharpnull#1{}
\def\aa{Astron. Astrophys.\ }

\null\voffset=+0.0pc  
\begin{document}

\slugcomment{\bf}
\slugcomment{Accepted to Ap.J.}

\title{A New Mechanism for Core-Collapse Supernova Explosions}

\author{A. Burrows\altaffilmark{1},  
E. Livne\altaffilmark{2},
L. Dessart\altaffilmark{1},
C.D. Ott\altaffilmark{3},
J. Murphy\altaffilmark{1}}
\altaffiltext{1}{Department of Astronomy and Steward Observatory, 
                 The University of Arizona, Tucson, AZ \ 85721;
                 burrows@as.arizona.edu,luc@as.arizona.edu,jmurphy@as.arizona.edu}
\altaffiltext{2}{Racah Institute of Physics, The Hebrew University,
Jerusalem, Israel; eli@frodo.fiz.huji.ac.il}
\altaffiltext{3}{Max-Planck-Institut f\"{u}r Gravitationsphysik,
Albert-Einstein-Institut, Golm/Potsdam, Germany; cott@aei.mpg.de}

\begin{abstract}

In this paper, we present a new mechanism for core-collapse supernova explosions
that relies upon acoustic power generated in the inner core as the driver. In our simulation
using an 11-M$_{\odot}$ progenitor, an advective-acoustic oscillation \`a la Foglizzo with a 
period of $\sim$25$-$30 milliseconds (ms) arises $\sim$200 ms after bounce.
Its growth saturates due to the generation of secondary shocks, and kinks in the resulting shock structure
funnel and regulate subsequent accretion onto the inner core.  However, this instability 
is not the primary agent of explosion.  Rather, it is the acoustic power generated 
early on in the inner turbulent region stirred by the accretion plumes,  
and most importantly, but later on, by the excitation and 
sonic damping of core g-mode oscillations.  An $\ell=1$ mode 
with a period of $\sim$3 ms grows at late times to
be prominent around $\sim$500 ms after bounce. The accreting
protoneutron star is a self-excited oscillator, 
``tuned" to the most easily excited core g-mode.  
The associated acoustic power seen in our 11-M$_{\odot}$ simulation 
is sufficient to drive the explosion $>$550 milliseconds after bounce.
The angular distribution of the emitted sound is fundamentally aspherical.
The sound pulses radiated from the core steepen into shock waves that
merge as they propagate into the outer mantle and deposit their energy 
and momentum with high efficiency.  The ultimate source of the acoustic 
power is the gravitational energy of infall and the core oscillation
acts like a transducer to convert this accretion energy into sound.
An advantage of the acoustic mechanism is that acoustic power does
not abate until accretion subsides, so that it is available as long
as it may be needed to explode the star. This suggests 
a natural means by which the supernova is self-regulating.

\end{abstract}


\section{Introduction}
\label{intro}

The essence of the mechanism of core-collapse supernovae must be the conversion
of a fraction of the reservoir of gravitational energy into the kinetic and internal
energy of the exploding mantle of the Chandrasekhar core whose instability inaugurates
core collapse.  A protoneutron star (PNS) is left which evolves into a cold neutron 
star most of the time, or, rarely, is the intermediate state 
along the way to the formation of a stellar-mass black hole through the 
general-relativistic instability \footnote{Direct formation of a black hole from a quasi-degenerate
Chandrasekhar core is not possible, since the inner homologous core
whose bounce halts collapse is out of sonic contact with the outer mantle.  The inner
core has a mass of but 0.5-0.7 M$_{\odot}$; it is only after sufficient mass
has accreted subsequent to bounce that the object might experience the 
general-relativistic instability (which acts on sub-millisecond timescales)
leading to collapse to a black hole.}.  Due predominantly to prodigious neutrino radiation 
(losses) over hundreds of milliseconds to seconds,
the PNS becomes more and more bound, on its way to achieving a total binding
energy near $\sim$3$\times$10$^{53}$ ergs, a mass-energy equivalent of $\sim$0.15 M$_{\odot}$.
This is far in excess of the $\sim$10$^{51}$ ergs of the supernova explosion,
so it has long been thought that 1) the mechanism of explosion is very inefficient,
and 2) there is so much energy in the neutrino emissions that if only a small
fraction were captured in the mantle it could be driven to explosion.  The latter
notion is the essence of the ``neutrino-driven" mechanism, first adumbrated
in its direct form by Colgate \& White (1966) and Arnett (1966) and in its delayed
form by Wilson (1985) and Bethe \& Wilson (1985). However, using the best 
input physics and numerics, it has been shown that in spherical symmetry 
both the prompt and the delayed-revitalization neutrino heating mechanisms   
fail (Rampp \& Janka 2000; Liebend\"{o}rfer et al. 2001; Thompson, Burrows, \& Pinto 2003).  
Moreover, the suggestion by Wilson (1985), Wilson \& Mayle (1988,1993), and Mayle \& Wilson (1988)
that ``neutron-finger" instabilities interior to or near the neutrinospheres 
can advect heat, and thereby boost the driving neutrino luminosities sufficiently to lead to
pseudo-spherical explosions, has been dealt a severe blow by the more detailed
analyses of Bruenn \& Dineva (1996), Bruenn, Raley, \& Mezzacappa (2004), and Dessart et al. (2005).
Finally, through two decades of simulations, the idea that on dynamical times the bounce itself could,
by direct piston action on the outer mantle, launch a successful shock wave
that does not stall into accretion has been refuted. Due to significant shock breakout neutrino losses
and nuclear photodissociation, the shock must stall.

Out of the ashes of the one-dimensional (1D), spherical models was born the idea that multi-dimensional
neutrino-driven convection could increase the efficiency of neutrino heating behind the 
stalled shock and lead to a delayed explosion (Herant et al. 1994; Burrows, Hayes, \& Fryxell 1995;
Janka \& M\"uller 1996; Fryer \& Warren 2002,2004). Such neutrino-driven convection
arises naturally and does increase the radius of the stalled shock and the 
size of the so-called ``gain region" external to the neutrinospheres where 
there is net neutrino energy deposition.  This is the current paradigm and models
that did not explode in 1D have indeed exploded in 2- and 3-D. However, all these previous successful
explosions were obtained in the context of simplified or gray treatments of neutrino transport 
and it has long been suspected that more sophisticated, multi-group, multi-angle
treatments of neutrinos in the multi-dimensional spatial context were needed.  In fact, though
implementing 2D neutrino transport in an approximate way, Rampp \& Janka (2002)
and Buras et al. (2003,2005) use 2D hydrodynamics and one of the best 
available multi-group co-moving-frame Boltzmann solvers and do not 
obtain explosions\footnote{unless they drop their many velocity-dependent terms, something 
they do not advise doing}.  Despite incorporating the best input physics, and coming tantalizingly close
to explosion, their simulations lead to fizzles, calling into question the neutrino-driven
mechanism itself.  Within 200-300 milliseconds of bounce, after growing out to 150-200 kilometers,
their shocks recede and are declared to be unsuccessful.

If the energy transfer from the core to the mantle necessary to 
explode the star is by neither direct hydrodynamics nor neutrino heating, what is
left?  How do core-collapse supernovae explode?  In this paper we propose a new alternative,
the generation in the core and the propagation into the mantle of strong sound waves.
Acoustic power is, potentially, an efficient means to transport energy and momentum into
the outer mantle to drive the supernova explosion.  Unlike neutrinos, sound is
almost 100\% absorbed in the matter, though some of the acoustic energy 
is reradiated by neutrinos.  As sound pulses propagate outward down the density gradient
they steepen into multiple shock waves that catch up to one another and merge; a shock wave is
almost a perfect black-body absorber of sound.  If sufficient sound is generated in the core,
it would be a natural vehicle for the gravitational energy of infall to be transferred
to the outer mantle and could be the key missing ingredient in the core-collapse 
explosion mechanism. Furthermore, periodic shocking due to multiple sound pulses
can lead naturally to entropies in the debris of hundreds of units, just what is required for 
r-process nucleosynthesis (Woosley \& Hoffman 1992; Woosley et al. 1994; Hoffman et al. 1996).

We believe we have identified a vigorous source for the necessary acoustic power:
the excitation and oscillation of core pulsation modes in the deep interior of the PNS.  Using
the 2D radiation/hydrodynamic code VULCAN/2D we have discovered that turbulence and 
anisotropic accretion in the inner 40$-$100 kilometers can excite and maintain vigorous core g-mode 
oscillations which decay by the radiation of sound.  The inner core acts as a transducer for
the conversion of accretion gravitational energy into acoustic power.   The associated acoustic power
seen in our simulations is sufficient to drive the explosion $>$550 milliseconds after bounce.

In \S\ref{approach}, we summarize the input physics, computational capabilities, 
and numerical approaches incorporated into the current version of our radiation/hydrodynamics 
code VULCAN/2D.  Then, in \S\ref{sim} we describe our baseline simulation results for
the 11 M$_{\odot}$ progenitor star on which we focus.  This section defines the various hydrodynamic
stages and events and explores the important, but secondary, role of the advective/acoustic or ``SASI" instability
highlighted by Foglizzo and collaborators (Foglizzo 2000,2001; Foglizzo \& Tagger (2002);
Foglizzo, Galletti, \& Ruffert 2005) and by Blondin, Mezzacappa, \& DeMarino (2003).
We also introduce the key role played at late times by acoustic power and the emergence 
of core oscillation modes and discuss some aspects of the 
components of the mechanism we envision.  Then, in \S\ref{acoustic}, we summarize  
the general advantages of acoustic driving, and follow this in \S\ref{core} with an investigation into  
the excitation of core oscillation modes, the associated general 
physics, and their acoustic damping.  After this, we summarize in \S\ref{neutrinos} the 
emerging role of neutrinos as important, but secondary, players in the
explosion mechanism we are presenting. Finally, in \S\ref{conclusion} we reiterate 
the central features of our new mechanism, explain why it was not discovered earlier, and speculate
on what now needs to be done both to test our hypothesis and to further advance 
our understanding of this central problem in theoretical astrophysics.

\section{Computational Approach and Input Physics}
\label{approach}

The code VULCAN/2D uses the hydrodynamic approach described in Livne (1993),
with the transport methods discussed in Livne et al. (2004) and Walder et al. (2005).
It is a Newtonian, 2D, multi-group, multi-angle radiation/hydrodynamics 
code with an Arbitrary-Lagrangian-Eulerian (ALE) structure (with remap) and a 
Multi-Group, Flux-Limited Diffusion (MGFLD) variant.  Velocity terms in the transport sector, such as 
Doppler shifts, are not included in the code, though advection is.  Note that the velocity
terms in Eulerian transport are different from the coresponding terms in the comoving frame
and that general statements about their relative importance are very frame-dependent\footnote{In particular,
even the {\it sign} of the effect can change.}.
We parallelize in energy groups using MPI and do not include redistribution by neutrino-electron scattering
in the 2D code.  Such redistribution and scattering are of modest import on infall, affecting
the trapped electron fraction (Y$_e$) and entropy ($S$) by $\sim$$10$\%, but are otherwise quite subdominate.
At a neutrino energy of 10 MeV, the neutrino-electron scattering cross section is $\sim$100 times
smaller than the dominant cross sections off nucleons.  Thompson, Burrows, \& Pinto (2003)
have devised and implemented an explicit scheme to handle redistribution that adds only
$\sim$$10-15$\% to the computational time and this approach will be incorporated in
future versions of VULCAN/2D.  Note that the attempt to handle the full energy/angle redistribution
problem implicitly has resulted in codes that are thereby slower by many {\it factors} (not percent),
severely inhibiting their use for explorations in supernova theory.

In this paper, due to its relative speed,  
we use the MGFLD implementation of VULCAN/2D. The flux limiter is
a vector version of the one found in Bruenn (1985).  The code can handle 
rotation, though the results reported here are for a non-rotating progenitor.  
In 2D, the calculations are axially/azimuthally symmetric,
and we use cylindrical coordinates ($r$ and $z$), but the grid points
themselves can be placed at arbitrary positions.  This allows us to employ
a Cartesian grid at the center (inner $\sim$20 kilometers) and transition
to a spherical grid further out.  The grid resolution is essentially
uniform everywhere within $\sim$20\,km.  A version of this grid structure is plotted
in Ott et al. (2004).  The Cartesian format in the interior allows us to avoid
the severe Courant problems encountered in 2D by all other 
groups employing grid-based codes due to the inner angular 
Courant limit and, thereby, to perform the calculations
in full 2D all the way to the center.  In many simulations to date, the inner core 
was calculated in 1D and grafted onto an outer region that was handled in 2D
(e.g., Burrows, Hayes, \& Fryxell 1995; Swesty \& Myra 2005ab), or
was excised completely (e.g., Janka
\& M\"{u}ller 1996; Blondin, Mezzacappa, \& DeMarino 2003;
Scheck et al. 2004).  In others, the 2D simulation was done on a $\sim$90$^{\circ}$ wedge,
with the inner 2 or 25 km done in 1D (Buras et al. 2005).
The gray SPH simulations of Herant et al. (1994) and Fryer \& Warren (2002,2004)
are an exception.  Originally, a major motivation for this global 2D  
feature was the self-consistent investigation of core translational motion and 
neutron star kicks.  However, as we will see, freeing the core has other advantages.

To determine the gravitational potential field we use a finite-difference approach to 
solve the Poisson equation.  In addition, we incorporate the 
gravitational force along the symmetry axis in an automatically
momentum-conserving fashion by writing it in divergence form. 
VULCAN/2D has an option to move the grid 
after bounce to maintain the best zoning under the core, whether
it moves or not, while at the same time tracking this core motion.  This feature 
ensures that the highest resolution is placed under most of the mass.  
Outside of the 
Cartesian mesh, the baseline calculation employs 121 angular zones
equally spaced over 180$^{\circ}$, and logarithmically allocates 162 radial shells
between $\sim$20\,km and the outer radius at 3800\,km\footnote{For future calculations,
the outer boundary radius should probably be positioned still further out.}.  

We employ the equation of state (EOS) of Shen et al. (1998), since it correctly 
incorporates alpha particles and is easily extended to lower densities.
%
The neutrino-matter interaction physics is taken from Thompson, Burrows, \& Pinto (2003)
and Burrows \& Thompson (2004).  The tables generated in T/$\rho$/Y$_e$/neutrino-species
space incorporate all relevant scattering, absorption, and emission processes. 
We follow separately the electron neutrino ($\nu_e$) and anti-electron neutrino ($\bar{\nu_e}$),
but for computational efficiency we lump the four remaining known neutrinos into ``$\nu_{\mu}$" 
bins in the standard fashion.  Our baseline model has 16 energy groups for each species,
distributed logarithmically from 1 to 200 MeV.  More energy groups would be better.
Due to extreme matter-suppression effects, we have not felt it necessary 
to incorporate the effects of neutrino oscillations (Strack \& Burrows 2005).

The instabilities that develop in the early stages of the post-bounce
phase are seeded by the slight perturbations introduced due to the 
non-orthogonal shape of the grid regions that effect the transition 
from the inner Cartesian grid to the outer spherical grid (see Fig. 
4 in Ott et al. 2004) and by noise at the part in $\sim$10$^6$ level 
in the EOS table interpolation.  Since the resulting turbules execute
more than twenty overturns during the initial phase of convective 
instability, and this convective phase reaches a quasi-steady state,
the initial conditions and the initial perturbations are completely 
lost in subsequent evolution. The seeds for the later shock instability 
are the non-linear convective structures that arise in the first 
post-bounce tens of milliseconds. Beyond these, we introduce no artificial
numerical perturbations.

\section{Simulation Results}
\label{sim}

To demonstrate and represent the various core-collapse and post-bounce phases we think are important,
we focus in this paper exclusively on results using the 11-M$_{\odot}$ 
progenitor without rotation from Woosley \& Weaver (1995),
with the zoning and physics packages referred to in \S\ref{approach}.  Progenitor 
dependences will be left to future work.  The calculations were done from $\sim$200
millseconds before bounce to $\sim$660 milliseconds after bounce, significantly longer 
than most (though not all) multi-D simulations in the literature \footnote{Typically, one complete
run will take four weeks on a state-of-the-art Beowulf cluster.}.  We have generated movies of 
the runs, which are available from the authors upon request. Here, we provide 
a sequence of stills depicting phenomena or transitions of relevance, along with 
analysis plots that help to clarify salient features.

\begin{figure*}
\includegraphics[width=8.5cm]{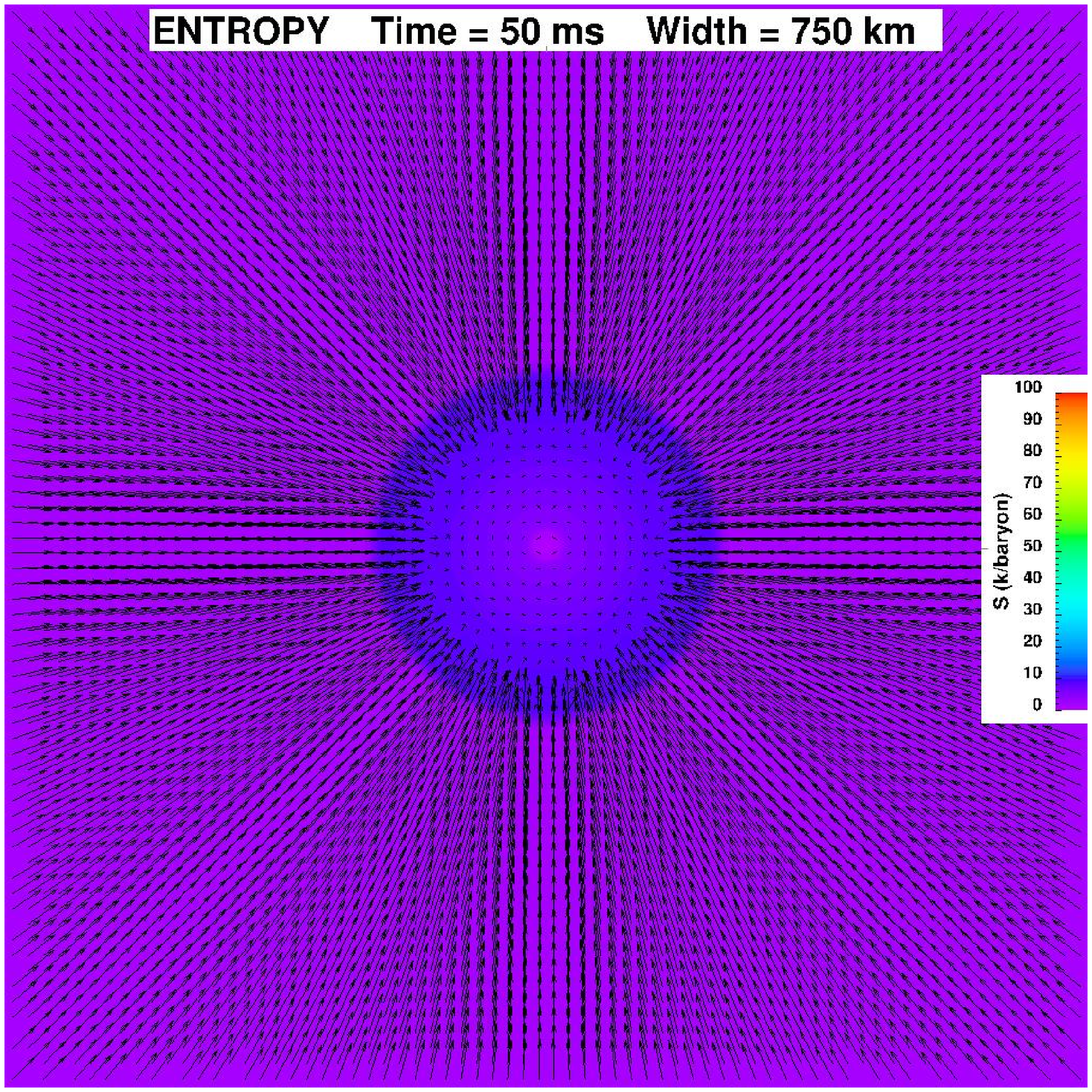}
\includegraphics[width=8.5cm]{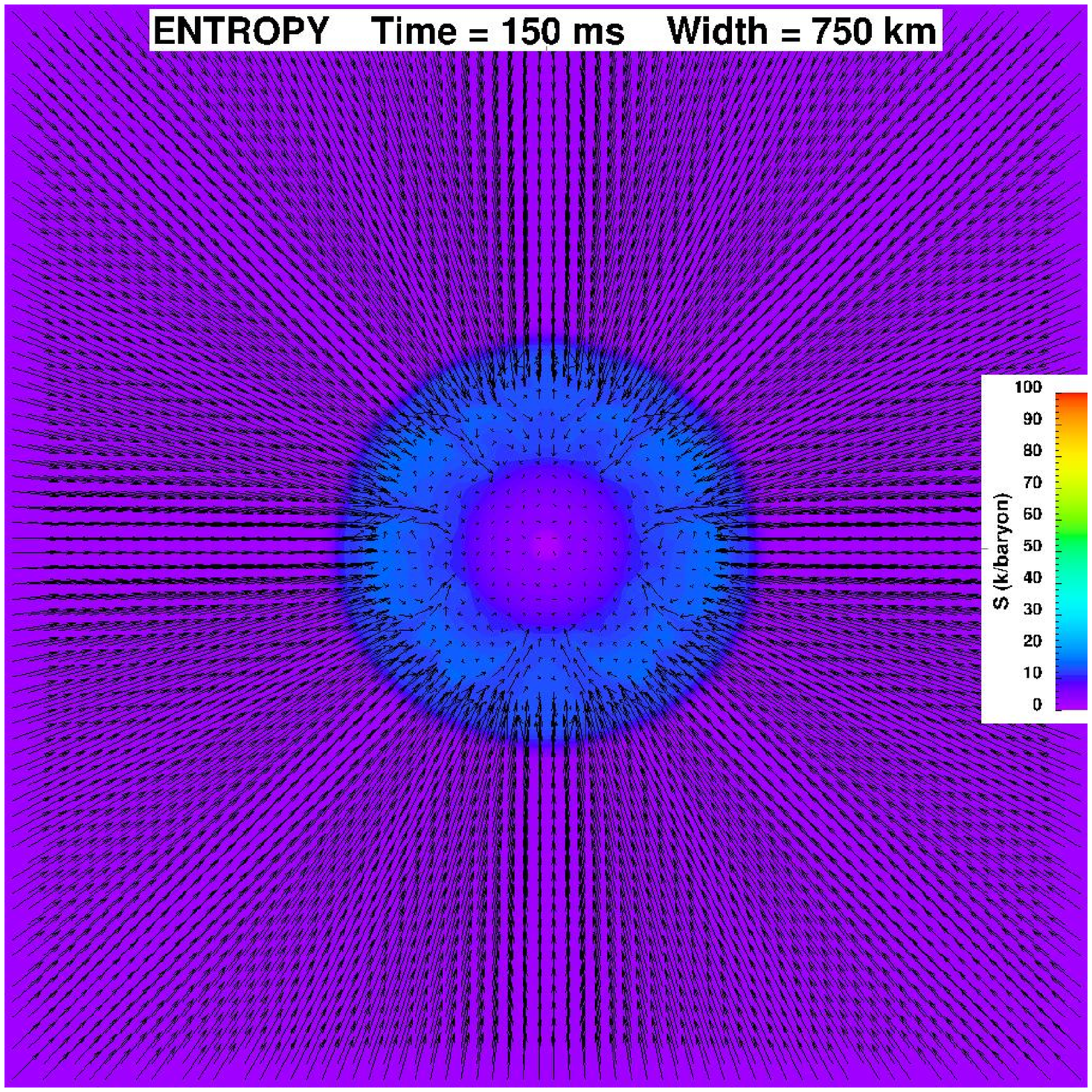}
\vspace{0.75cm}
\includegraphics[width=8.5cm]{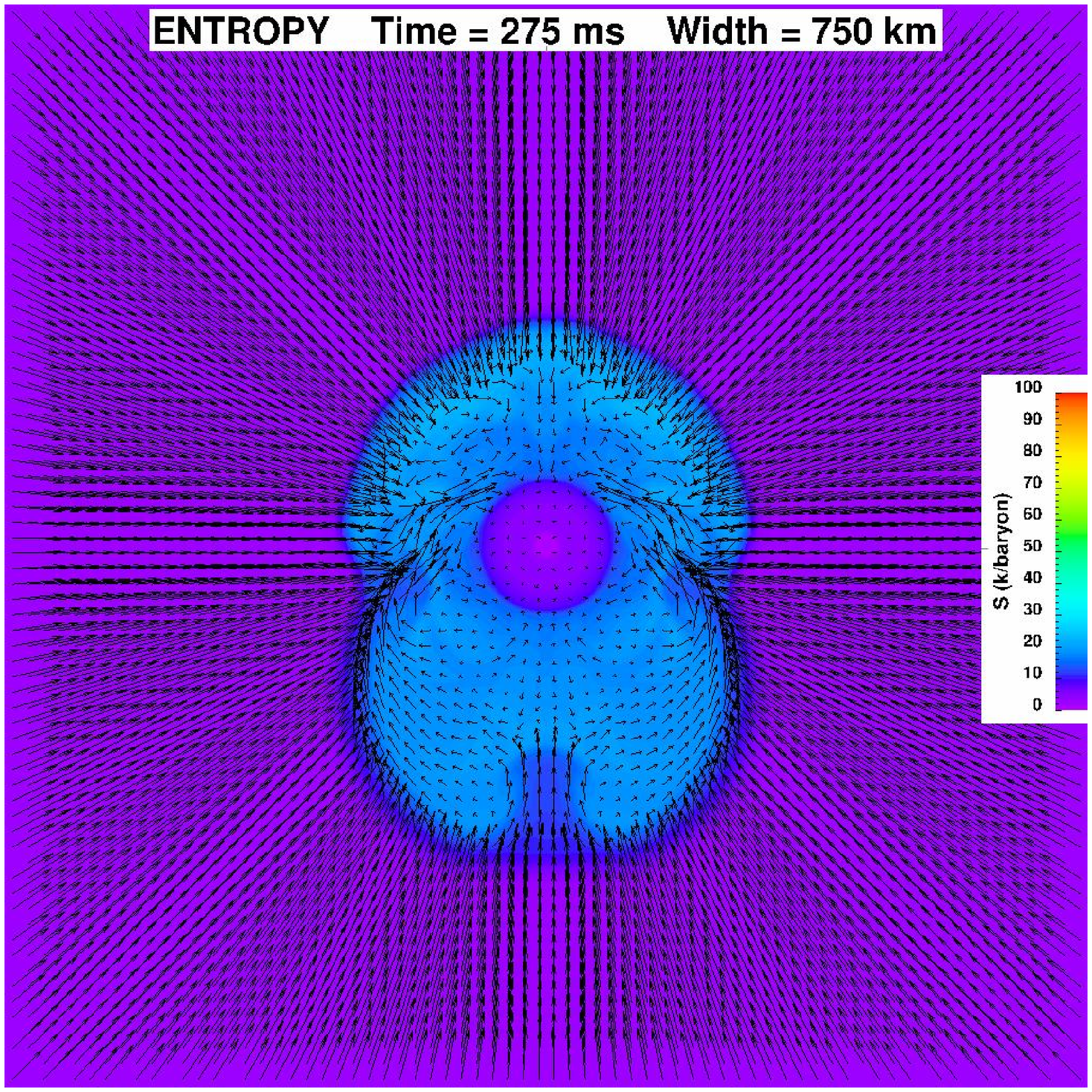}
\includegraphics[width=8.5cm]{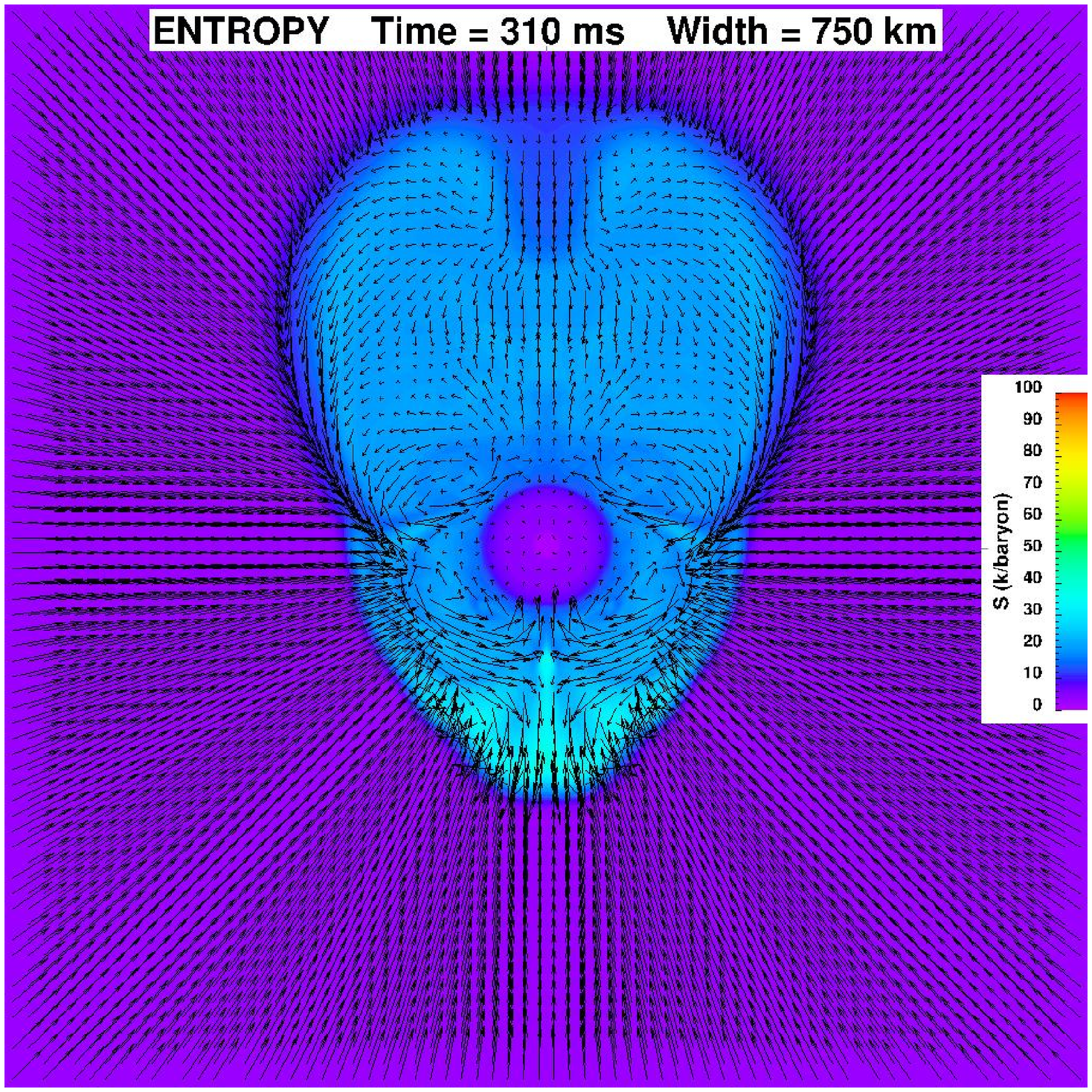}
\caption{Colormap stills of the entropy, taken at 50 (top left), 150 (top right),
275 (bottom left), and 310\,ms (bottom right) past core bounce, with velocity
vectors overplotted.  Here ``Width" refers to the diameter through the middle;
the radius through the middle is 375 kilometers.
Note that on this figure, as well as on Fig. \ref{fig_ent_intermed}, 
for ease of comparison between panels the same 
colormap is used. It extends up to 100 units (red), above which
it saturates (See text for discussion).  These calculations have been done 
for a full 180$^{\circ}$ and the axis of symmetry is vertical.
\label{fig_ent_early}}
\end{figure*}

\begin{figure*}
\includegraphics[width=8.5cm]{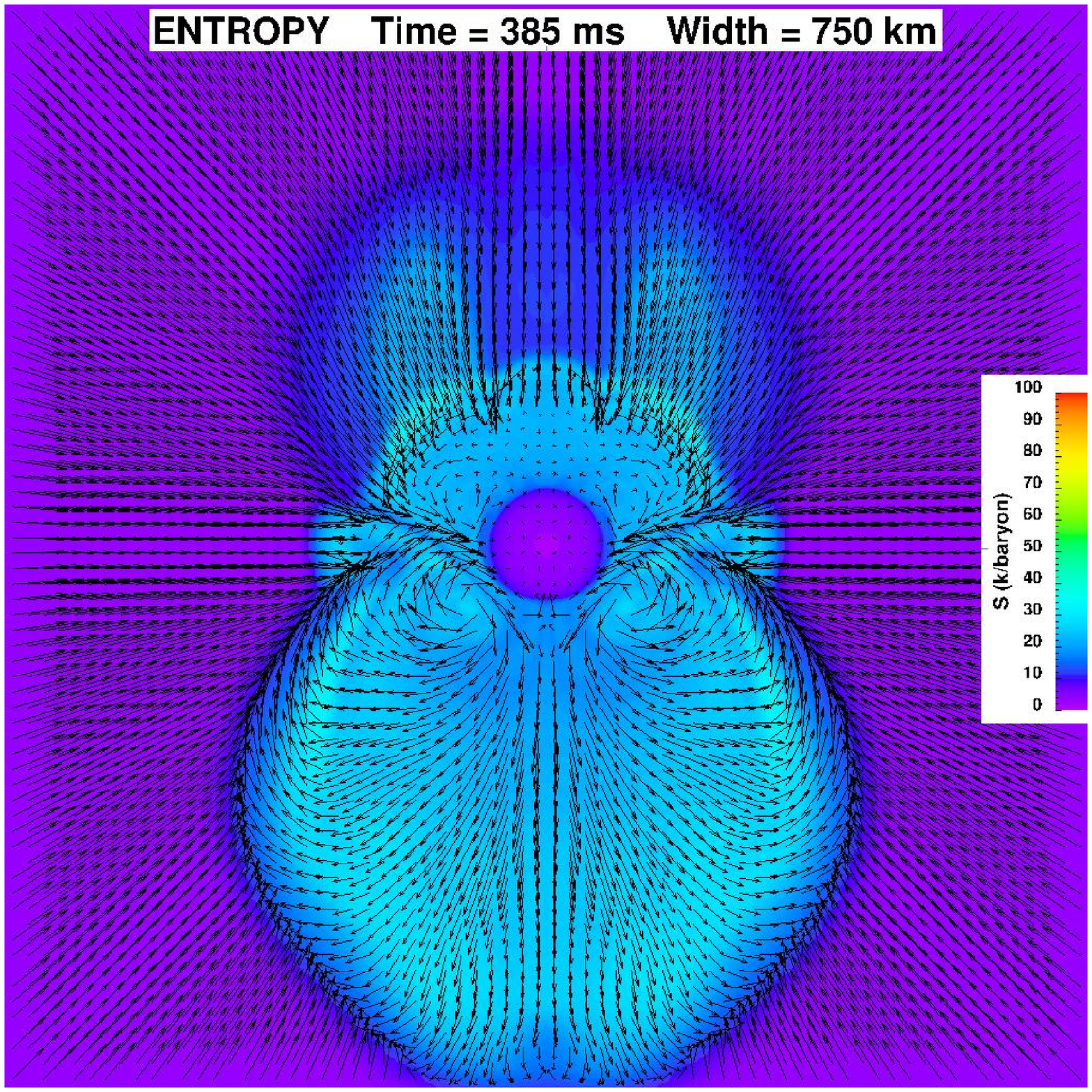}
\includegraphics[width=8.5cm]{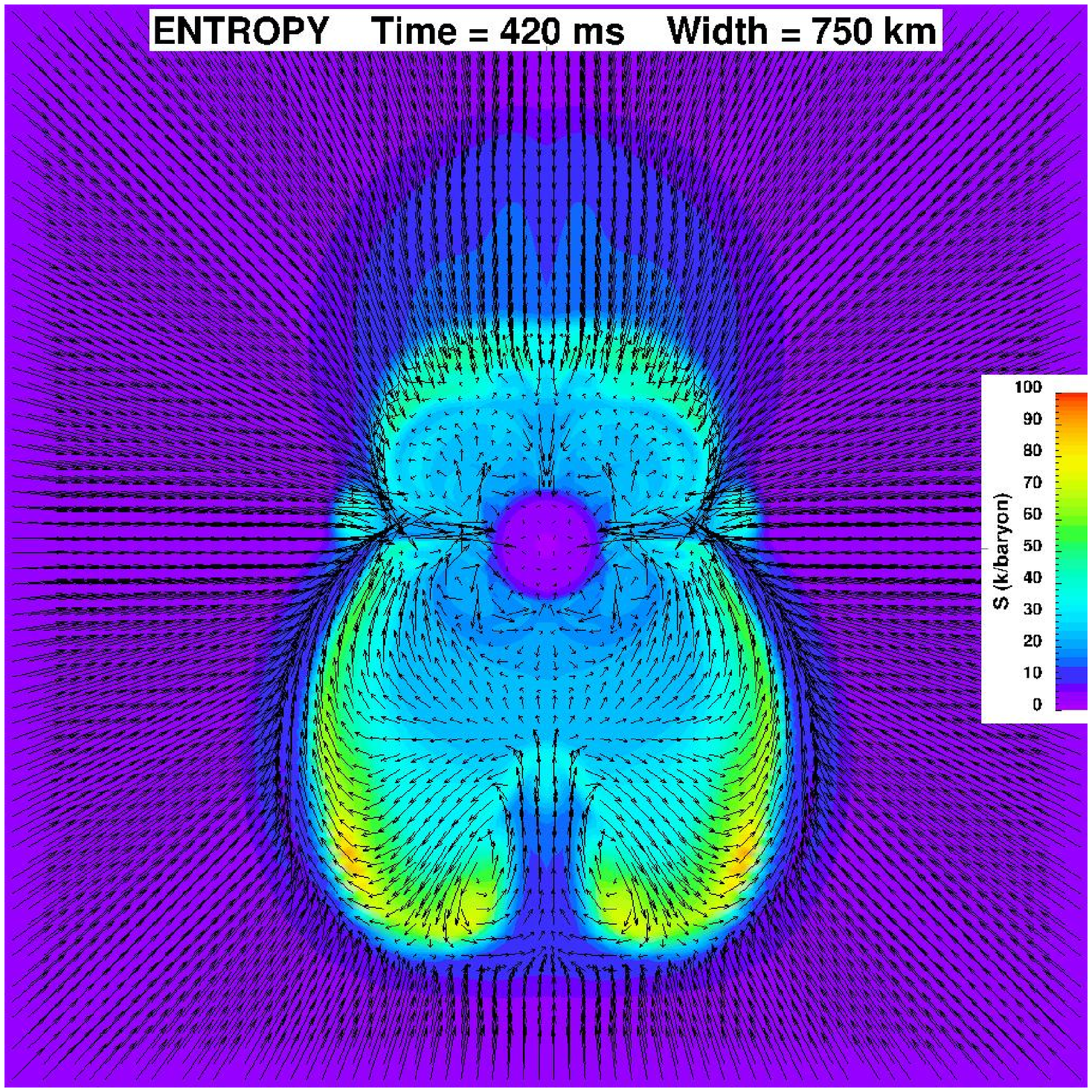}
\vspace{0.75cm}
\includegraphics[width=8.5cm]{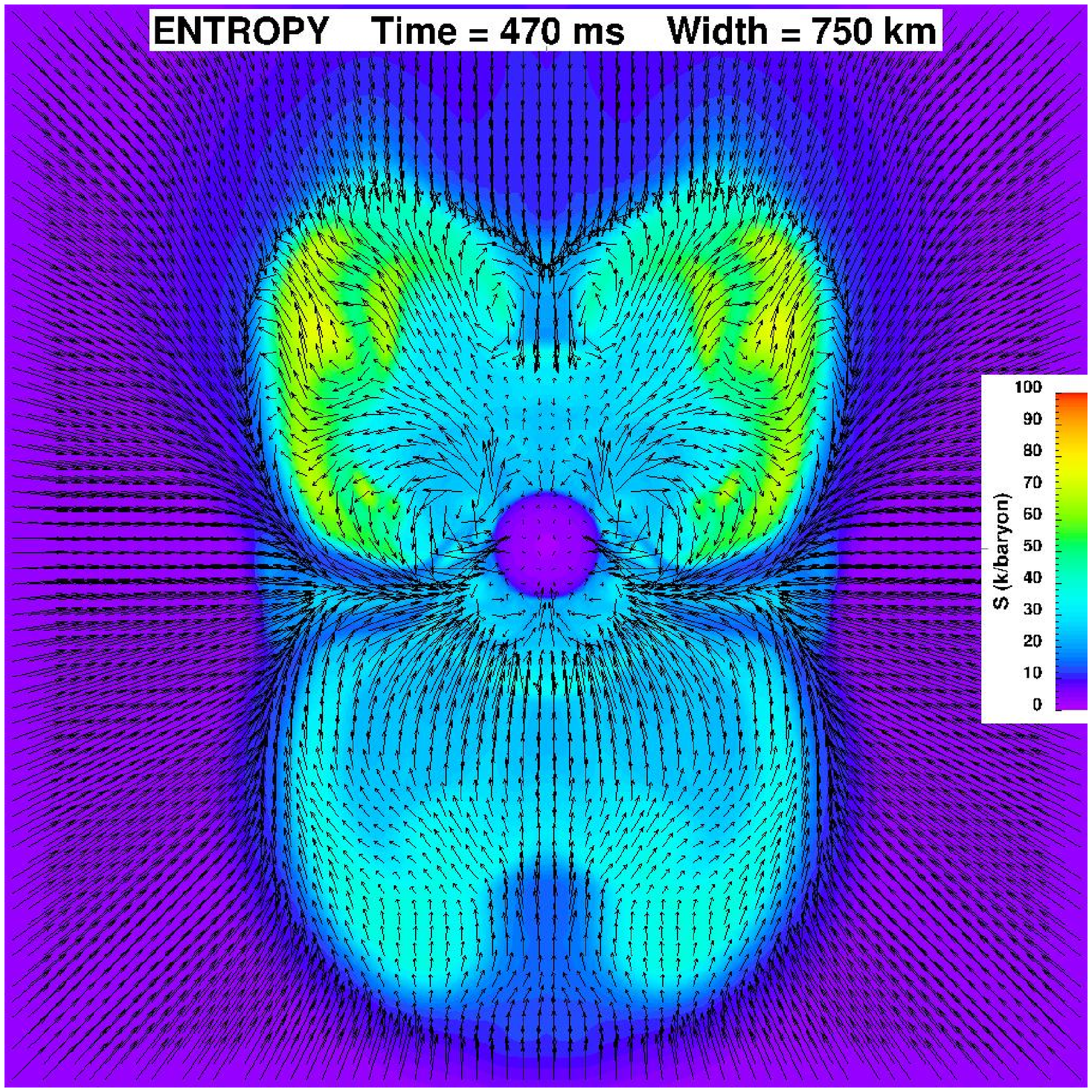}
\includegraphics[width=8.5cm]{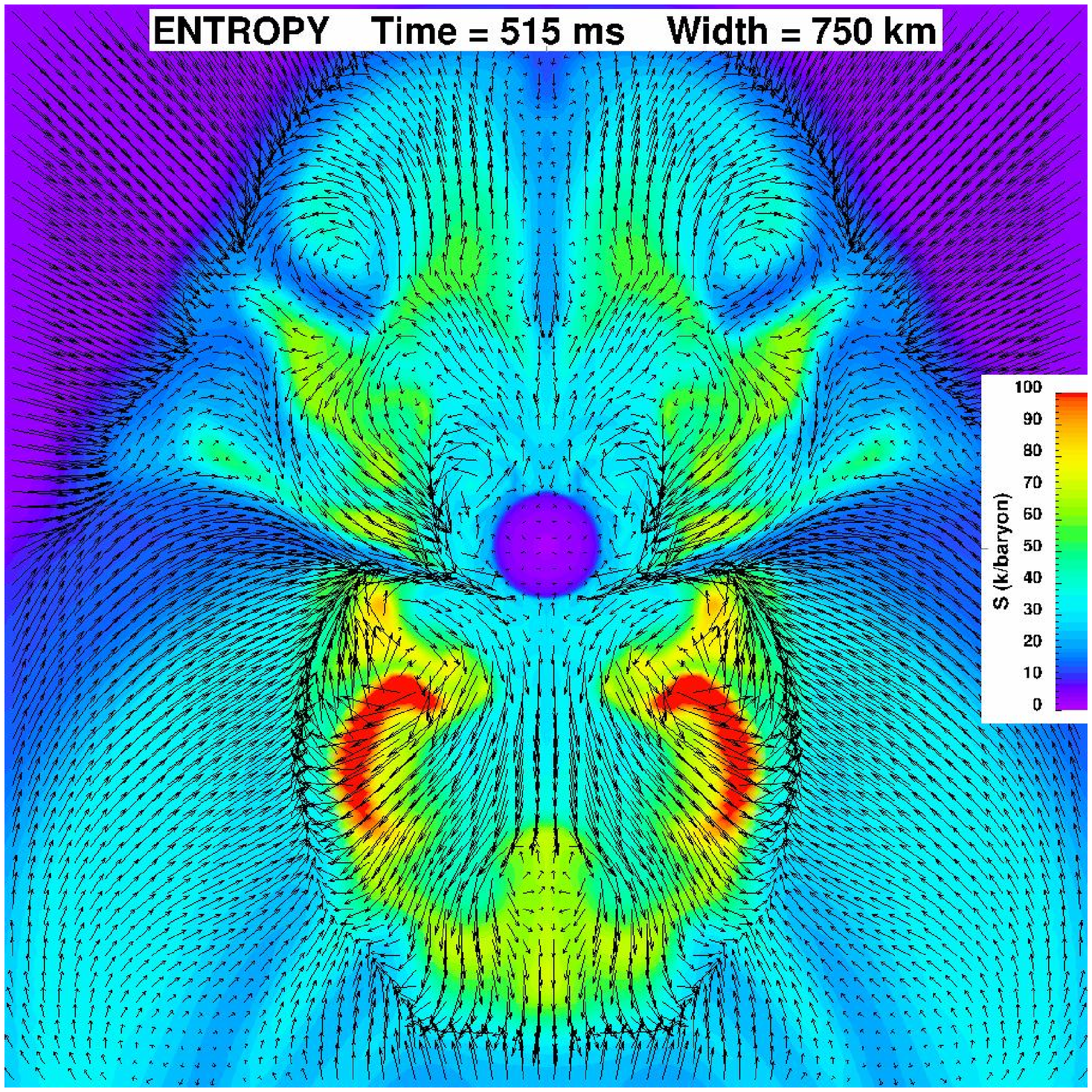}
\caption{
Same as Fig.~\ref{fig_ent_early}, but this time showing the entropy
at 385 (top left), 420 (top right), 470 (bottom left), and 515\,ms
(bottom right) past core bounce (See text for discussion).   
}
\label{fig_ent_intermed}
\end{figure*}

\begin{figure*}
\includegraphics[width=8.5cm]{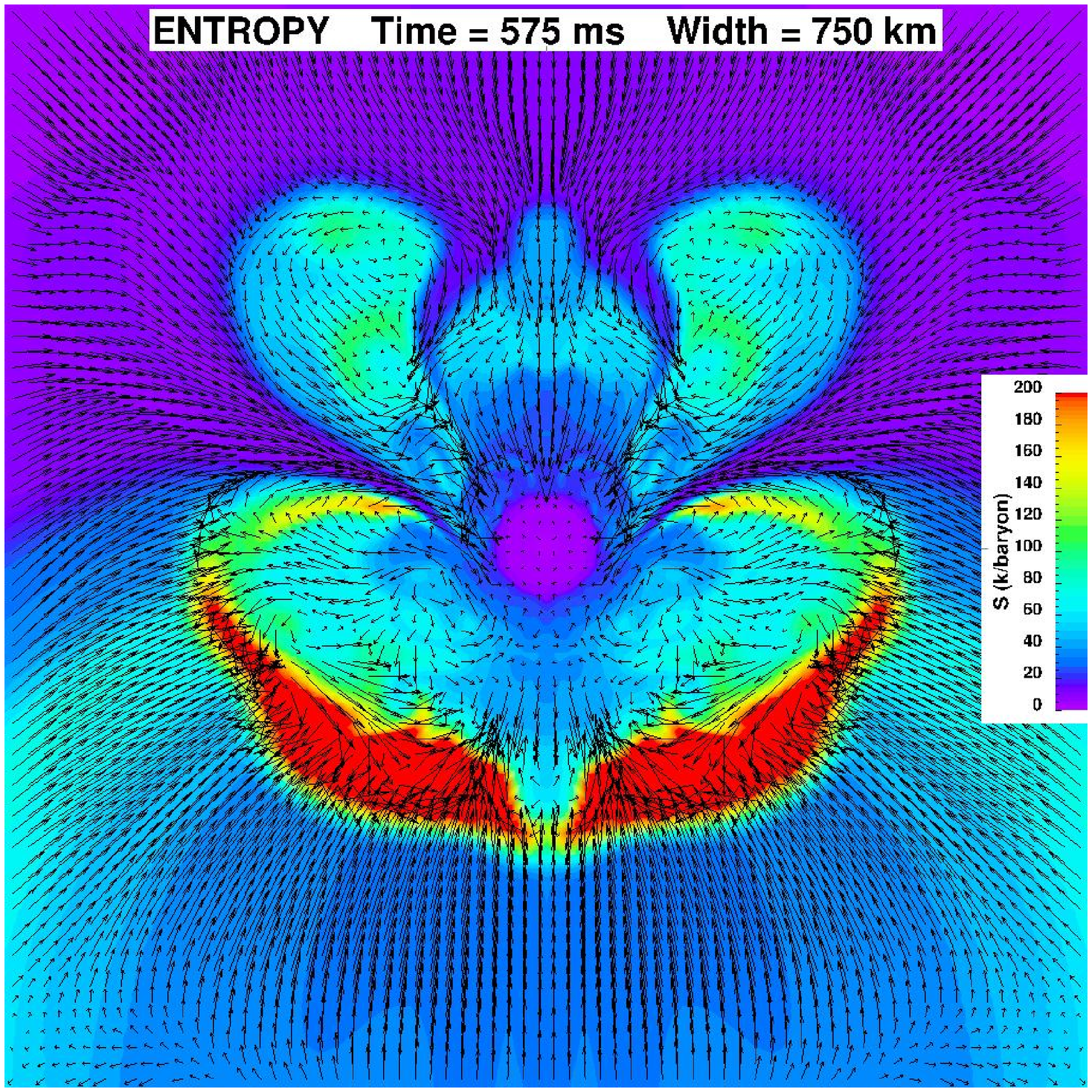}
\includegraphics[width=8.5cm]{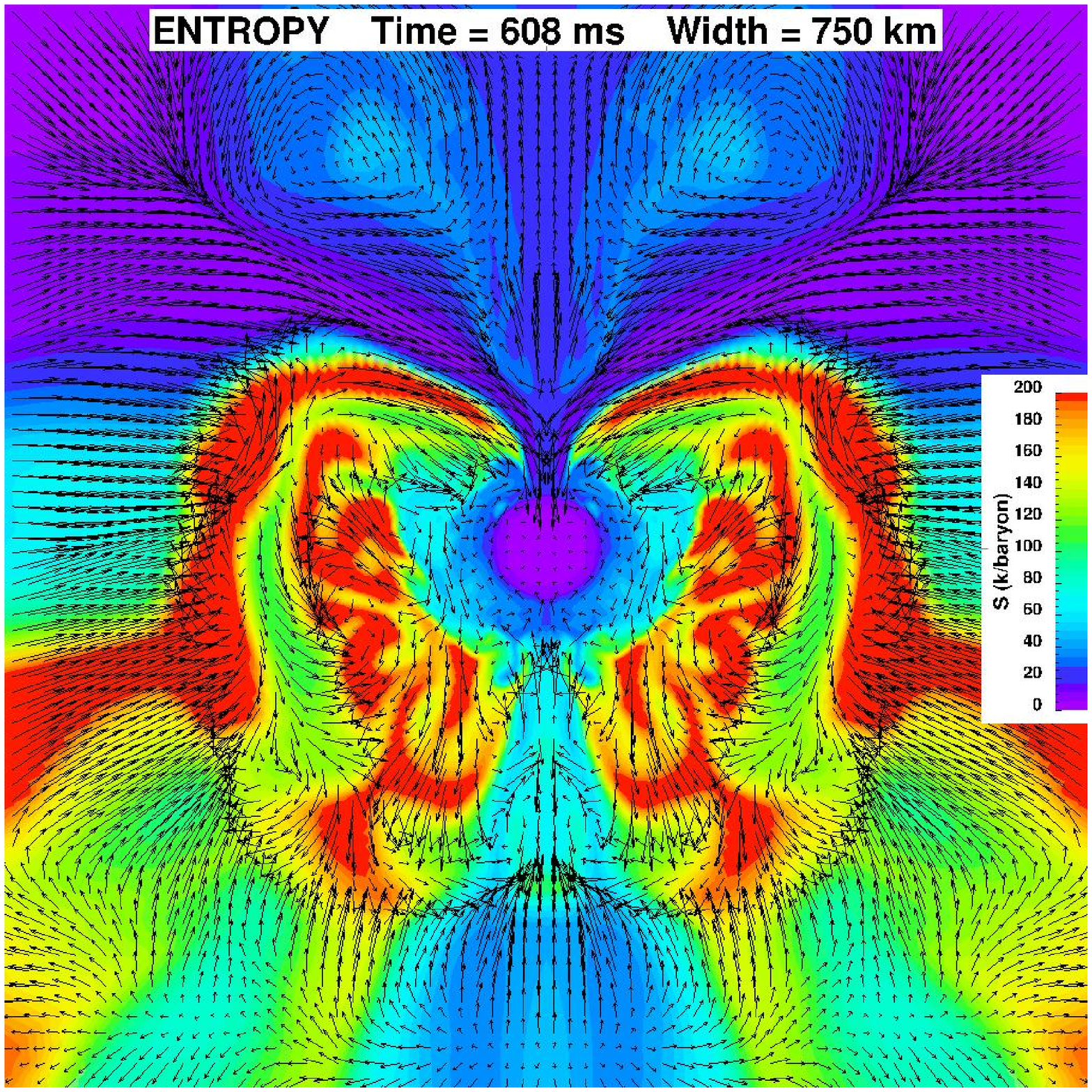}
\vspace{0.75cm}
\caption{
Same as Fig.~\ref{fig_ent_early}, but this time showing the entropy
at 575 (left) and 608\,ms (right) past core bounce.  Note the acoustic
waves emanating from the core, most easily seen 
in the velocity vector field.  The color map extends
to entropies of 200 (red), and then saturates for entropies
beyond 200.  The low-entropy accretion streams that are exciting the core g-mode
are clearly seen (see text for discussion).
}
\label{fig_ent_late}
\end{figure*}

Figures \ref{fig_ent_early} through \ref{fig_ent_late} are an evolutionary sequence
of the entropy distribution for the core region 750 kilometers on a side,
with velocity vectors superposed to map out the evolving and complicated 
flow fields.  The sequence includes snapshots at times of 50, 150, 275, 
310, 385, 420, 470, 515, 575, and 608 milliseconds after bounce.  In the 
last two stills, the supernova has clearly started to explode.

\begin{figure*}
\plotone{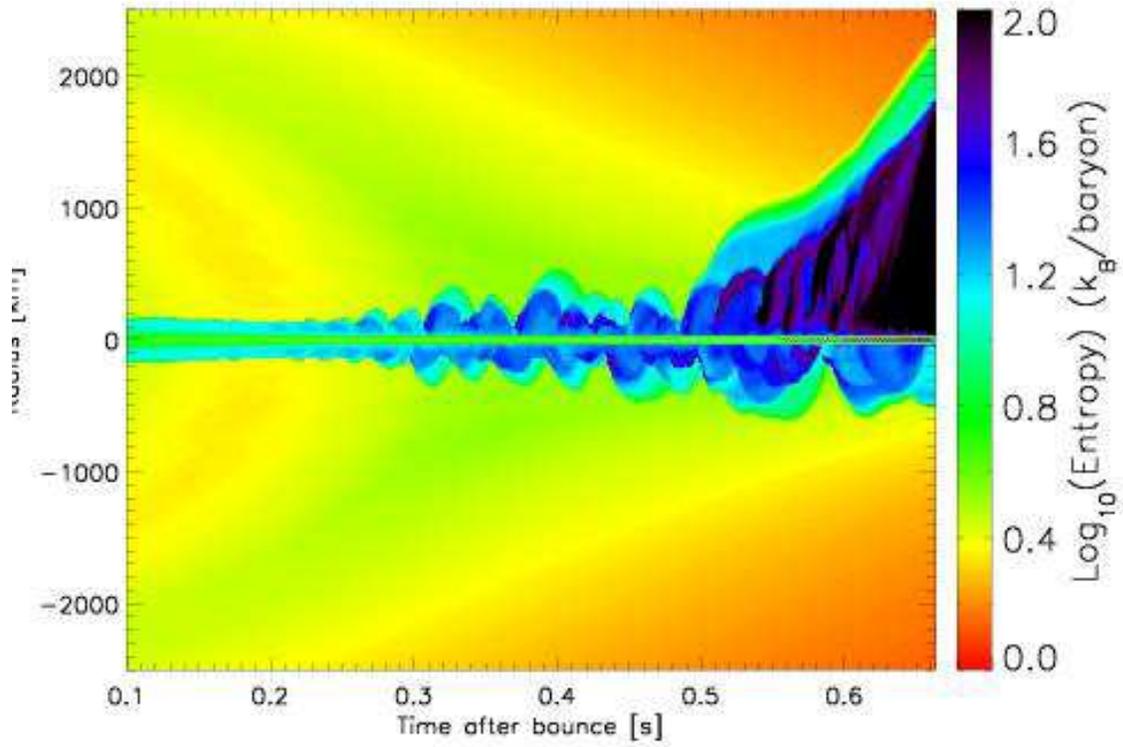}
\caption{Time evolution, from 100\,ms to $\sim$660\,ms after core bounce,
of the entropy (logarithmic scale) along the axis of
symmetry, i.e., at $r=0$ or latitudes $\pm$90$^{\rm o}$, and covering the inner
2500 kilometers of the grid.  An entropy ceiling of 100 has been adopted.
}
\label{ent}
\end{figure*}

\begin{figure*}
\plotone{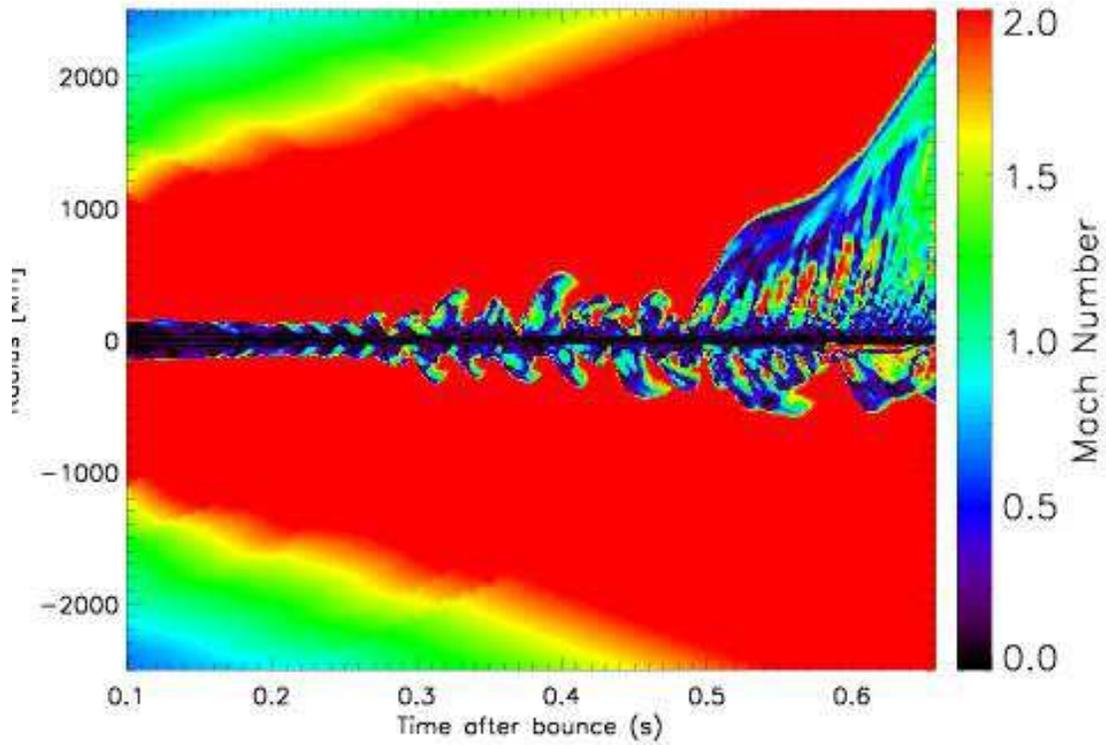}
\caption{Time evolution, from 100\,ms to $\sim$660\,ms after bounce, of the
Mach number along the axis of symmetry, i.e., at $r=0$ or latitudes
$\pm$90$^{\rm o}$, and covering the inner
2500 kilometers of the grid. Note how clearly the multiple 
secondary shocks emerging at late times can be discerned 
and that the displayed Mach number saturates at 
a value of two.}
\label{mach2}
\end{figure*}

In the sequences depicted in Figs. \ref{fig_ent_early}, \ref{fig_ent_intermed}, \& \ref{fig_ent_late}, we 
don't show the collapse phase because it is canonical and spherical.  At $\sim$50 ms
after bounce the shock has stalled, is roughly spherical, and is at a spherical radius ($R$)
of $\sim$115 km.  Neutrino-driven convection has begun in the region $\sim$50 km
wide interior to the shock wave.  Despite the advection correction to the 
standard convective instability criterion pointed out by Foglizzo, Scheck, \& Janka (2005),
we find that standard neutrino-driven convection at this stage has not been suppressed,
though there is evidence of a weak $\ell=1$ pulsation similar to what has
been identified by Foglizzo and collaborators and by Blondin, Mezzacappa, 
\& DeMarino (2003) (Figs. \ref{ent} and \ref{mach2}).
By $\sim$150 ms, the average shock radius has reached $\sim$150 km, and the convection
is encompassing the region down to $R \sim 75$ km.
In the full angular region of 180$^{\circ}$, we see 
5-6 dominant turbules (eddies) with angular scales of $\sim$30-35$^{\circ}$.  Near $\sim$200
ms, the average shock radius has receded back to $\sim$110 km  
(Figs. \ref{ent} and \ref{mach2}).  If shock recession
had been our criterion for failure, we might have stopped the calculation here.
However, at around $\sim$200 ms, the shock is beginning to wobble up and down
perceptibly in an $\ell=1$ mode with a period near 25-30 ms and a $\Delta{R}/R$ near 25\%. 
At $\sim$250 ms, the $\Delta{R}/R$ is 
approaching $\sim$50\% and the up-down asymmetry is quite pronounced.
The growth time for the shock anisotropy varies, but is near 50-100 ms.
We identify this early quasi-periodic oscillation with the advective-acoustic instability
and the ``standing accretion shock instability" (SASI) suggested by Foglizzo (2001,2002),
Foglizzo \& Tagger (2000), Foglizzo, Galletti, \& Ruffert (2005), and
by Blondin, Mezzacappa, \& DeMarino (2003), referred to from now 
on as the ``shock instability."  

By $\sim$300-350 ms, $\Delta{R}/R$
has grown to a factor of two and the wobble is taking on a more vigorous character.
However, by $\sim$275 ms it has clearly entered the nonlinear regime.  In particular,
when one side is executing its outward oscillation, the material in its outer extent
reverses its flow direction out of phase with the material further in.  The result
is the generation, where these two regions collide, of a secondary 
shock wave; the flow now has nested shock waves.  The
creation of secondary shocks serves to saturate the amplitude of the shock
oscillation, which nevertheless continues.  Furthermore, the corrugation of the outer
shock, where there is a kink in the shock normal, propagates in latitude in   
both hemispheres and collides at the poles, first at 0$^{\circ}$ then at 180$^{\circ}$.
In part, this shock focussing is an artifact of the 2D nature of the simulation,
but similar effects are seen in 3D pure hydro simulations (J. Blondin, 
private communication) and are expected from generic considerations.
Each time the shock kinks collide, entropy and pressure ($p$) are generated and this pressure pulse moves at
the speed of sound to the opposite side.  Importantly, at the shock intersection kinks,
clearly seen in Figs. \ref{fig_ent_early} and \ref{fig_ent_intermed} at 310 and 385 ms, the accretion through
the outer shock is channeled and penetrates in lower entropy streams into the interior 
and onto the core.  This phenomenon was previously seen by, among others, Herant et al. (1994), Burrows, 
Hayes, \& Fryxell (1995), Fryer \& Warren (2002), Rampp \& Janka 
(2002), Scheck et al. (2004), Janka et al. (2005ab), and Buras et al. (2005).  
Though $\Delta{R}/R$ can reach factors of 3 or greater, this ratio always eventually 
decreases, if temporarily, and stabilizes.  The 
angle-averaged outer shock radius does not increase monotonically.  For instance,
though at $\sim$385 ms, one lobe of the outer shock reaches $\sim$400 km, as Fig. \ref{fig_ent_intermed}
indicates, at $\sim$420 ms the average radius has receded temporarily to $\sim$250 km. 
The 2D hydro calculations of Blondin, Mezzacappa, \& DeMarino (2003) that identified this instability
were done without neutrino physics, and with an artifical inner boundary; they were done
without electron capture, neutrino heating, neutrino cooling, and the core.  
Nevertheless, with neutrino physics and the core fully included, we verify the 
existence of this shock instability, first promoted by Foglizzo (2000,2001).  

However, we have discovered significant and important differences with the work of
Blondin, Mezzacappa, \& DeMarino (2003) and with the expectations from the work of Foglizzo
and collaborators.  First, the shock instability does not lead to an increase in the average shock
radius in the first 150-200 ms after bounce.  Due to electron capture 
at the shock and to the quasi-hydrostatic sinking 
of the inner core caused by steady neutronization and cooling, an ``$\ell=0$" component
is not in evidence.  Shock asphericities with some $\ell=1$ component are generated, but mostly by 
the turbulent motions of neutrino-driven convection during this early phase.  It is only
after the core has settled, the electron capture rates at the shock have decreased due to the gradual
decrease in the density of the accreted material, and there is no net energy loss
behind the shock in the gain region that the advective-acoustic behavior identified by 
Foglizzo is clearly manifest.  This delay could not have been captured in the calculations of Blondin et al. (2003).
Second, when the shock instability can finally be identified, the matter is already 
convecting nonlinearly due to neutrino heating.  Hence, a linear growth analysis may be
inaccurate, or, at the very least, the seed perturbations for the shock-instability are  
the turbules and plumes of neutrino-driven convection.  Third, the nonlinear phase of
the shock instability brings with it secondary shock waves 
and the shock oscillation becomes saturated and non-periodic.  By itself, the shock instability
is not leading to explosion and the average radius of the outer shock ceases to increase,
though the wobble and top-bottom asymmetry can still be extreme.  Fourth, within 
the first 100-150 ms of the shock instability its behavior not only ceases to be 
periodic with a clear oscillation period, but is also not a simple normal mode. 
A new phenomenon arises. The increasing vigor of the turbulence in the interior outside
of the core ($< 100$ km), in part generated by the fluctuations and ``wagging" over 
the core's surface of the lower-entropy accretion streams stirring this inner region,
begins to generate traveling acoustic waves, sound that propagates outward, 
with initial periods $\le$5$-$10 ms determined by the
turbule overturn times in the interior and the timescale for the heads of the accretion
plumes to traverse the latitudes of the PNS surface.  Such acoustic power is a new feature 
in supernova theory.  This is not the acoustic component of the 
``advective-acoustic" instability, which is a normal mode with a period near the sound
travel time across the shocked region of $\sim$25-30 milliseconds.  Rather, it is the 
{\it propagating} acoustic flux whose source begins as the turbulent energy in the roiling interior,
driven by aspherical accretion of matter and entropy onto the core.  This is analogous to
the ``forced-turbulence" source for the acoustic flux studied by Goldreich \& 
Kumar (1988,1990) in the context of solar convection. These authors  
showed that the efficiency of sound generation in a driven turbulent region 
increases roughly as the Mach number ($M$) cubed \footnote{Furthermore, they showed that
the peak in the acoustic frequency spectrum is near $H/v$, which is close to what we witness
at this stage in our simulation, where $H$ is the characteristic
turbule size and $v$ is the characteristic turbulent speed.  
However, another characteristic driving timescale is the
period over which the accretion streams ``dance"/``flap" over the surface 
of the PNS. This timescale is a bit longer than $H/v$ until 
the core g-mode oscillation starts to dominate,
at which point the flapping all but ceases.}.  
The Mach numbers seen in our simulation are on average 
steadily rising, evolving in the inner $\sim$100 km from $\sim$0.1 to as high as $\sim$0.9. 
This turbulence is initially fed by the grossly aspherical accretion flows 
associated with the shock instability and the dancing accretion streams regulated by it.  

However, when the acoustic flux becomes
appreciable, it significantly modifies the shock instability.  In our calculation, over a period from
$\sim$350 to $\sim$450 ms, the motions of the shock and accretion flow transition from being
dominated by the shock instability to being dominated by the core's acoustic flux.  The
propagating sound waves issuing from the core modify the outer flow to such an extent that the shock
motion is no longer even quasi-periodic and the momentum and energy flux of the sound start to 
determine the flow pattern.  In particular, the acoustic flux is refracted and reflected by the 
accretion streams and emerges very anisotropically, and more on one side of the core than another.
At the same time, the accretion streams are pushed by the momentum of the acoustic flux more and more onto
the other side of the core, whichever that happens to be.   The positions of the accretion streams and the 
average direction of the acoustic flux become anti-correlated.  Figure \ref{fig_ent_late}  
clearly shows the low-entropy accretion streams and their positioning, as well as the discontinuities in the velocity
vector field that mark the emerging sound pulses.  In fact, in propagating outward down the
density gradient, the sound pulses steepen into shock waves, subsequent pulses catching 
up with and reinforcing previous pulses.  From $\sim$400 ms after bounce, sound waves 
make themselves felt more and more and the resultant nested shock waves become a 
feature of the early supernova phenomenon.

By 500 ms after bounce, the acoustic power from the core is quite pronounced and 
is starting to power outflow and the beginnings of an explosion.  
Figure \ref{ent} depicts the temporal evolution after 100 ms of the entropy profile along the poles.
The position of the outer shock, the early decay of its radius, the wobble due 
to the shock instability after $\sim$250 ms, the 
fact that the outer shock radii at 0$^{\circ}$ and 180$^{\circ}$ (the top and the bottom) are roughly out of phase 
until $\sim$500 ms, the entropization due to successive pulses at late times,
and the onset of explosion are all clearly seen.  
Figure \ref{mach2} depicts the evolution of the Mach number profile, also along the axis
of symmetry and as a function of time. This figure recapitulates
Fig. \ref{ent}, showing the shock oscillations and the onset of explosion, while
also showing the growth of the Mach number with time and the secondary shocks at later times.
Note that despite clear evidence on this figure for a periodicity in the Mach number of 
the accreted matter and, hence, in the flow from $\sim$100 ms to $\sim$250 ms after bounce, the 
average shock radius does not increase due to a shock instability until afterwards. 
By $\sim$500 ms, one lobe of the outer shock is near 1000 km and by $\sim$660 ms 
it has reached $\sim$2300 km. The accretion streams are now perennially on one side 
of the core and the acoustic flux, while mostly emanating from the other side is 
radiating in all directions, except up the accretion streams themselves.  Figure 
\ref{fig_ent_late} depicts this clearly, though at $\sim$600 ms 
some matter is still infalling, particularly the lower-entropy streams. 
The entropy of some of the shocked matter can now be hundreds of units, due mostly to
progressive shock heating by the steepening sound pulses, not the neutrinos.
Starting around $\sim$550 ms, as the explosion commences, entropization due to these 
multiple shocks is clearly seen in Fig. \ref{ent}.

After $\sim$500 milliseconds, and certainly by 550 ms, the sonic power, though 
its early source was the turbulence around the core region, is driven mostly 
by a core oscillation that is being excited by the violent 
accretion streams. This oscillation can be seen as early as $\sim$350 ms 
after bounce and has a period near $\sim$3 milliseconds, 
very much smaller than the sound-travel-time in the shocked cavity.
It is predominantly an $\ell=1$ g-mode of the inner $\sim$40 kilometers 
that has grown strong over a period of $\sim$100 ms to  
reach nonlinear amplitudes by $\sim$500 ms.  This mode would have been suppressed had we excised the inner core,
not performed the calculations over the full 180$^{\circ}$,
or performed the simulations in 1D interior to some convenient radius and is discussed 
in \S\ref{core}.  The core oscillation is driven by the 
energy in the accretion streams and by the turbulence around the core, both of which ultimately derive
their energy from the gravitational energy of infall.  The oscillation is damped by sound
waves that emerge out of the core.  These sound waves steepen 
into shock waves, and, by dint of their momentum and energy flux
\footnote{the latter, minus that fraction radiated away by neutrinos},
``ignite" the supernova explosion. The core oscillation is acting like a transducer for 
the conversion of the gravitational energy of infall into radiating acoustic 
power and at the later stages is a far more important source of acoustic power than the
inner turbulence. Moreover, from $\sim$400 ms to $\sim$660 ms the efficiency 
for the conversion of accretion power into sound power is increasing.  As 
long as the accretion continues during this phase, the core 
oscillation seems to be driven and the sound is emitted.  After the explosion has 
progressed sufficiently and accretion subsides, the core oscillation decays and the sound source abates.
It seems that as long as the acoustic power due to core oscillation is needed to drive the explosion, it is available.
If the neutrino mechanism does not abort this scenario by inaugurating an earlier explosion,
this may be a natural self-regulating mechanism for the supernova phenomenon.

\section{Acoustic Waves as the Critical Power and Momentum Source}
\label{acoustic}

Hydrodynamic, neutrino, convective, viscous (Thompson, Quataert,
\& Burrows 2005), and magnetic (Akiyama et al. 2003) mechanisms for driving
core-collapse supernova explosions have all been proposed and investigated.
Acoustic power is (or would be) a new paradigm.  If the neutrino mechanism obtains,
it needs to inaugurate explosion early, perhaps in the first 200-400 milliseconds
after bounce.  This is because the driving neutrino luminosity,
the absorbing mantle mass, and the optical depth of this mass to the 
emerging neutrinos, are decreasing.  The result is that the neutrino energy deposited 
in the gain region is inexorably diminishing.  In the simulations we present
here, we do not see a neutrino-driven explosion.  For all the
most detailed simulations by the Garching group (cf. Buras et al. 2005), except 
for their model s15Gio$\_$32.a, which they exploded artificially 
(see also, Janka, Buras, \& Rampp 2003), this is also the case.
This does not mean that the neutrino mechanism does not obtain.  Indeed, 1) the LANL group
(cf. Fryer \& Warren 2002,2004) does get neutrino-driven explosions using its gray SPH
code, 2) many calculations to date may have critical flaws, and 3) accretion-induced collapse 
may well explode by the neutrino mechanism rather easily.

\begin{figure*}
\plotone{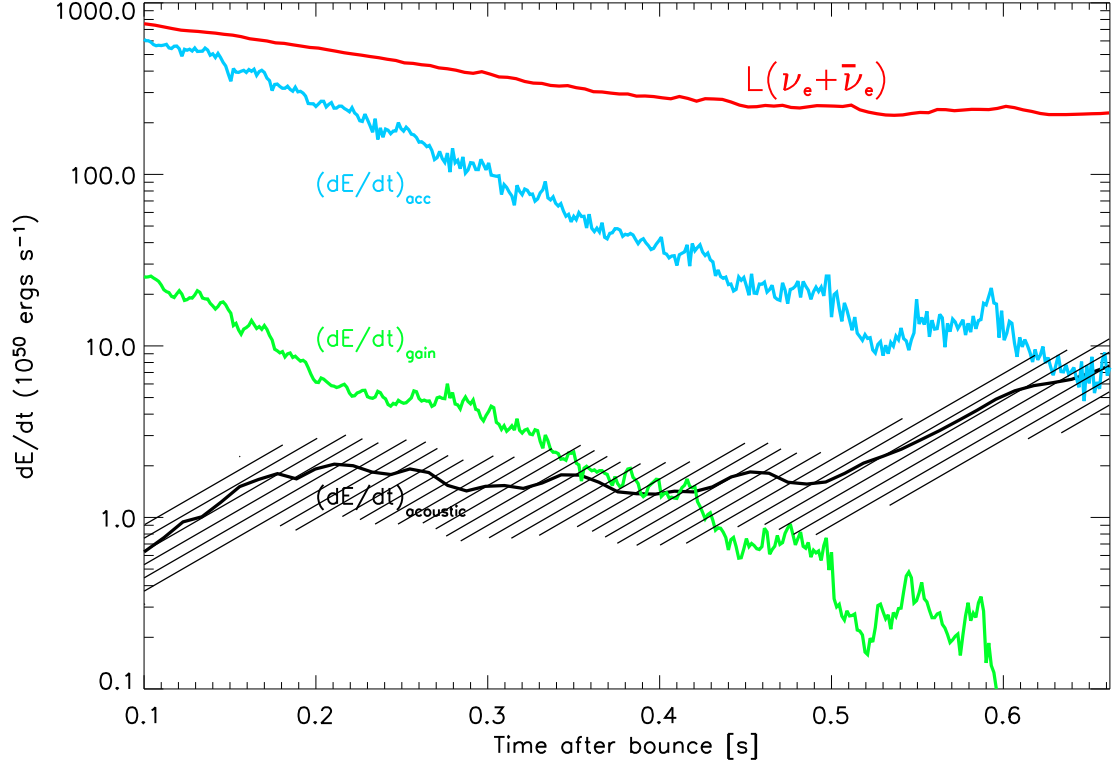}
\caption{
Time evolution, after core bounce, of the sum of the $\nu_e$ and
$\bar{\nu}_e$ neutrino luminosities ($L({\nu_e}+{\bar{\nu}}_e)$, red line);
the gravitational energy of the accreted material (blue) through a
radius $R=$35\,km, defined as $(dE/dt)_{\rm acc} = \frac{G M_R}{R}$$\dot{M}_R$,
where $M_R$ is the mass interior to R and $\dot{M}_R$ the mass accretion rate
through R; the net energy deposited by neutrinos in the gain region (green), defined as
$(dE/dt)_{\rm gain}= \int_{\Omega} \varepsilon(R,\theta) dm(R,\theta)$,
where $\Omega$ is the gain region, $\varepsilon(R,\theta)$ is the net
neutrino heating rate, and $m(R,\theta)$ is the mass of the
cell at $(R,\theta)$; and the acoustic power ($(dE/dt)_{\rm acoustic}$) radiated by the core 
oscillation.  This is approximated here by the ratio $E_{g}/{\tau_E}$, 
where $E_{g}$ is the total g-mode energy interior to a 
radius of 40\,km and $\tau_E$ is the decay time of the core pulsation, 
taken as 120\,ms. The default acoustic power (solid black line)  
threads through a $\pm$50\% swath that represents our estimate of the current ambiguity
in extracting the numerical acoustic power from the simulation.}  
\label{lum}
\end{figure*}

However, there are certain virtues to acoustic driving that bear mentioning.
First is that while the acoustic luminosity is much smaller than the neutrino
luminosity, almost all of the sound is absorbed in the mantle matter. At late
times in our simulation, less than a percent of the 
$\nu_e$ and $\bar{\nu}_e$ neutrino luminosity is absorbed. 
This amounts to an neutrino absorption power of $\le$$10^{50}$ erg s$^{-1}$, 
compared with an estimated core acoustic power at the end of our calculation  
near $\sim$${10}^{51}$ erg s$^{-1}$.  Figure \ref{lum} compares
the $\nu_e$ and $\bar{\nu}_e$ neutrino luminosities, the integrated neutrino
heating rate (power) in the gain region, the gravitational accretion
power ($\dot{E}_{acc} \sim \frac{GM}{R}\dot{M}$), and an estimate of the
acoustic power due to core g-mode oscillations.  Due in part to the very complicated flow
patterns and sonic refraction and reflection, it is difficult to determine precisely this 
total acoustic power.  However, with a simulation-motivated estimate
of $\sim$120 ms for the acoustic damping e-folding time ($\tau_E$) 
of the $\ell=1$ core g-mode \footnote{One can use the FWHM of the power 
spectrum of the core pressure fluctuations around the 3$-$ms mode or the early growth rate
of the core oscillation kinetic energy to get a handle on this quantity.} and our calculated 
energy ($E_g$) in this normal mode as a function of time, we can estimate this late-time acoustic
power ($\sim E_g/\tau_E$) and have included this estimate in Fig. \ref{lum}.  There is a time, in 
our calculation near $\sim$350$-$400 ms, when the absorbed neutrino 
power goes below the rising acoustic power.  

Second, sound
carries not only energy, but momentum, and this factor seems to be important in our
simulations.  The momentum flux for sound with the same energy flux as neutrino radiation
is larger by the ratio of the speed of light to the speed of sound, which in the
inner mantle regions is as much as a factor of ten.   Third, acoustic power propagates
from where it is generated to where it is needed; it fulfills the central requirement
of a core-collapse supernova mechanism that it involve energy transfer from the bound
interior PNS to the outer exploding mantle.  If the acoustic power is large 
enough, it is the ideal transfer agent.  Fourth, the acoustic source 
seems to grow just when the neutrino luminosity is ebbing and, importantly, it 
continues until explosion ensues.  Fifth, the successive merger of trains of sound waves 
that steepen into shocks provides a non-neutrino way to entropize some of the matter
and naturally achieve r-process conditions.

\section{The Excitation and Damping of Core Oscillations}
\label{core}

\begin{figure*}
\plotone{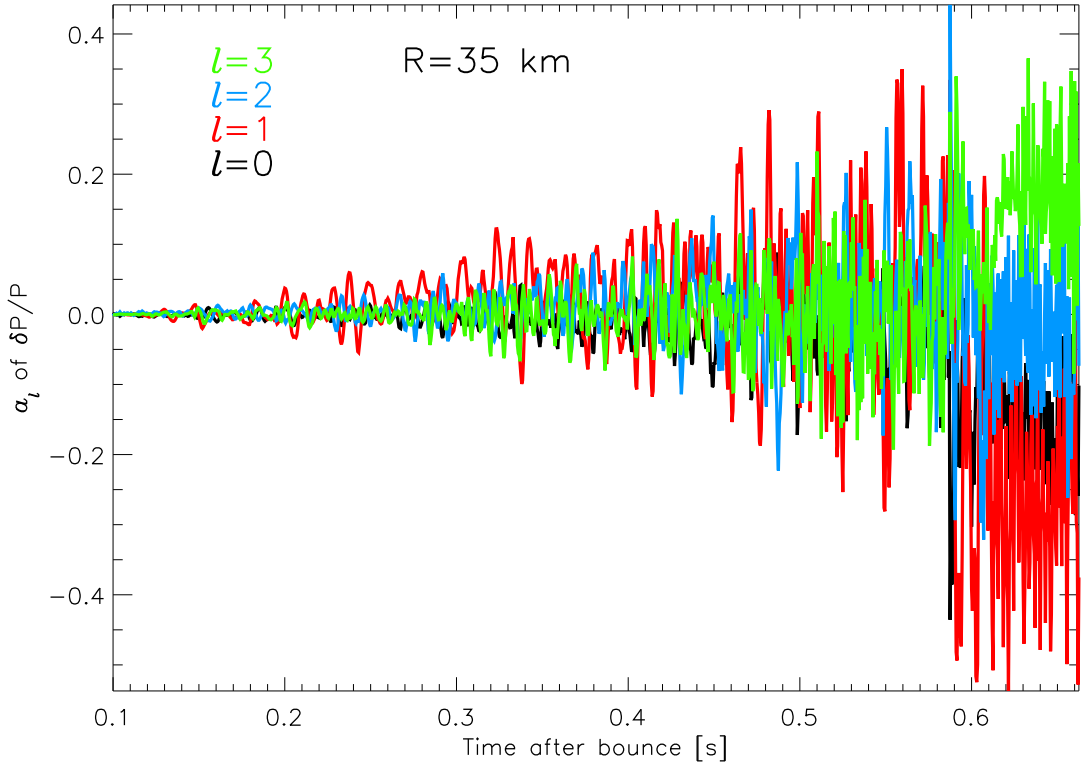}
\caption{Time evolution of the spherical-harmonic coefficients for the fractional pressure variation for
modes $\ell=$0 (black), 1 (red), 2 (blue), and 3 (green) at a radius $R = 35$\,km,
given by $a_{\ell} =  2 \pi \int_0^{\pi} d\theta \sin\theta Y_{\ell}^0(\theta)
(P(R,\theta)-<P(R,\theta)>_\theta) / <P(R,\theta)>_\theta$. Notice that despite the fact
that the $\ell=$1 mode looms large, the $\ell=$2 and $\ell=$3 modes are also in evidence.
The $\ell=$2 (harmonic) mode will result in a distinctive signature 
in gravitational radiation detectors, initially at a frequency near $\sim$675 Hz. 
This frequency is likely to be different (higher) when general 
relativity is included (Ferrari et al. 2003).
}
\label{press}
\end{figure*}

Within $\sim$200-300 ms of bounce, the inner core has reached 
a total baryon mass above $\sim$1.3 M$_{\odot}$.  Exterior to a radius
of $\sim$40 km, the density profile is falling off precipitously.
The entropy in this inner core is of order unity (in units of Boltzmann's constant per baryon), 
and the outer Y$_e$ profile has a steep negative gradient.  The neutrinospheres reside
in this low Y$_e$ regime at radii of $\sim$30$-$40 km.  Surrounding this inner core 
is a high-entropy accreting, turbulent mantle, whose entropy and Mach numbers are 
increasing with time. At $\sim$200 ms, the total accretion
rate is $\sim$0.1 M$_{\odot}$ s$^{-1}$ and is decreasing, but the temporal 
and spatial variations in the accretion fluxes and the ram pressures of the 
infalling plumes that eventually settle onto the inner core are quite large.   
Figure \ref{press} shows the time evolution of the spherical 
harmonic weighting coefficients, $a_{\ell}$, of a decomposition into Y$_{{\ell}m}$s of 
the fractional fluctuations of the pressure field at a radius 
of 35 km for $\ell=0,1,2,$ and 3\footnote{$\Delta{p(\theta)} = <p>\sum_{\ell} a_l Y_{{\ell}0}(\theta)$}.  
We see in Fig. \ref{press} the emergence of a pronounced $\ell=1$ component,
whose modulation up to $\sim$600 ms is at the period of the outer shock instability ($\sim$25$-$30 ms).

This situation is perfect for the excitation of the normal modes of oscillation of the inner core 
that are analogous to classical stellar pulsations (Goldreich \& Kelley 1977).  
Stars can execute p-modes, g-modes, and/or f-modes. The dominant restoring force 
for p-modes is pressure and that for g-modes is gravity. The 
f-modes generally have no radial nodes and are fundamentally of 
mixed type, though all realistic pulsational modes have some mixed 
character.  Importantly, all the prominent modes excited in our 
simulations are significant admixtures of both g-type and p-type character.  We 
have performed an analytic  modal analysis of the PNS core for its structure at 
$\sim$500 ms after bounce and display the derived periods in Fig. \ref{modal}.
The p-mode (green dots) and g-mode (red stars) branches for each spherical 
harmonic angular eigenfunction (Y$_{{\ell}m}$) are shown 
here as a function of $\ell$\footnote{Without rotation, we expect the modes to be degenerate in $m$.}.
Given an $\ell$, for predominantly g-modes the period increases with number 
of radial nodes and for predominantly p-modes the period decreases with number of radial nodes,
the latter as expected for sound waves in a cavity.  Generally, the shorter period
(higher frequency) oscillation modes are predominantly p-modes and those at longer periods (lower frequency) are 
predominantly g-modes and for all modal types the period is a decreasing function of $\ell$.
The radial eigenfunctions of p-modes peak on the outside, while those of 
g-modes peak on the inside.  At the boundary between the p-modes and the g-modes
on Fig. \ref{modal}, the modes are quite mixed, so that some of those modes might
be assigned differently by others, for instance as f-modes.

The periods for most of the modes depicted in Fig. \ref{modal} are shorter than the characteristic
times of pressure fluctuations and turbule overturn in the turbulent outer regions.  Moreover, the dynamical
times in the core-bounding region from 50 to 100 kilometers are longer than those in the core itself.
In addition, though excitation by a neutrino $\kappa$-mechanism has yet to be investigated,
one can show that the high-frequency p-modes would damp so quickly by sound emission
that they can not be excited to significant amplitude. However, as Goldreich \& Kumar (1990)
have shown, the efficiency for turbulence to excite g-modes is linear in Mach number, $M$.
As $M$ grows, the fraction of the turbulent power around the core that is pumped into 
g-modes increases.  Furthermore, the g-modes have the longest periods among the modes
that could possibly be excited, and are, hence, the most likely to be resonantly excited
by convective turbules and the waving accretion streams. 

Therefore, we would expect a whole spectrum of core modes to be excited, with the g-modes eventually
dominating.  A variety of $\ell$s above $0$ should be represented, but the fact that the accretion
is channeled into a small number of plumes/streams (generally, in our 2D calculations, one per hemisphere 
at late times) means that lower values of $\ell$ should prevail. This is what we see in our 2D simulations.
After $\sim$450 ms, over a period of $\sim$100 ms, the $\ell=1$ dipole mode with predominantly g-type
character at a period near $\sim$3 milliseconds emerges to significant amplitude, reaching 
Lagrangian displacements of $\sim$3 km.  This core mode has a radial node near 6$-$10 km, 
though it has a few nodes of p-type character further out.  This mode corresponds
to the analytic mode circled in Fig. \ref{modal}\footnote{There are no $\ell=0$
g-modes, and no g-modes without nodes, since this would not conserve momentum.  Note that the 
$\ell=1$ p-mode without a radial node would be a zero-frequency core translation, a ``kick."}. 
It damps by the emission of acoustic power, sound waves, with an oscillator $Q$ value ($=\omega\tau_E$, where
$\tau_E$ is the e-folding time for energy loss) near $\sim$200.  In its nonlinear phase after $\sim$580 ms,
the total pulsation energy is $\sim$$10^{50}$ ergs and the radiated acoustic power is  
near $\sim$$10^{51}$ erg s$^{-1}$. Figure \ref{fftp} depicts the time evolution of the spectrum 
of pressure fluctuations at a radius of 30 km.  On this frequency-time plot one sees the emergence of
the 3$-$ms core g-mode oscillation to dominate after $\sim$450 ms, as well as a number of other 
periodicities (modes) of lesser strength.  For instance, the mode seen in Fig. \ref{fftp} 
near $\sim$675 Hz is an $\ell=2$ harmonic.

Though the radiation pattern we actually observe is quite anisotropic and variable, and
the sound is severely refracted and diverted by the complicated accretion/turbulent flow patterns,
to gain insight one can estimate the dipole acoustic power ($\dot{E}^{\ell=1}_s$) radiated 
by a sphere of radius $R$ executing periodic linear 
translational motion (wobbling) of amplitude $\Delta R$ with period $P$ 
in a gas of density $\rho$ and sound speed $c_s$.
This is given by (Landau \& Lifshitz 1959):
\begin{equation}
\dot{E}^{\ell=1}_s = \frac{2\pi}{3}\rho{c_s}R^2u_0^2 \frac{(\kappa{R})^4}{(4 + (\kappa{R})^4)}\\
\label{eq:1}
\end{equation}
$$
\sim 4\times{10}^{51} {\rm erg\,} {\rm s}^{-1} \rho_{12}\Bigl(\frac{c_s}{3\times{10}^9 
{\rm cm\, s}^{-1}}\Bigr)\Bigl(\frac{R}{30\, {\rm km}}\Bigr)^2 
\Bigl(\frac{\Delta{R}}{3\, {\rm km}}\Bigr)^2\Bigl(\frac{3\,{\rm ms}}{P}\Bigr)^2 \, ,
$$
where $\rho_{12} = \rho/10^{12}\, {\rm g\, cm}^{-3}$, $\kappa$ ($= \omega/c_s$) 
is the wave number of the sound generated, $\omega$ is the angular
frequency of the oscillation ($=2\pi/P$), and $u_0 = \omega\Delta{R}$.  The numbers given in
eq. (\ref{eq:1}) are representative quantities for the problem at hand and the impedance term 
($(\kappa{R})^4/(4 + (\kappa{R})^4)$) has been set to 1/5.
Since in reality there are complicated density and sound speed profiles, 
the application of eq. (\ref{eq:1}) is no more than a very crude approximation.
Nevertheless, we see from simple analytics that even for small values of $\Delta{R}$
the acoustic power radiated by core oscillations can be quite
large, and can be of importance in the supernova context.  
At high oscillation frequencies and surrounding pressures, even small amplitude oscillations
generate competitive acoustic radiation.  This is what we see in the 
more detailed simulations, in which we derive acoustic powers due to core oscillations 
of $\sim$${10}^{51}$ erg s$^{-1}$.  Almost all of this power is absorbed in the mantle
and is potentially available to the explosion.  This is to be compared (see Fig. \ref{lum}) with the accretion power
at late times of:
\begin{equation}
\dot{E}_{acc} = \frac{GM}{R}\dot{M} \sim 10^{52}\, {\rm erg\, s}^{-1} 
\Bigl(\frac{M}{1.4 M_{\odot}}\Bigr)\Bigl(\frac{\dot{M}}{0.1 M_{\odot} {\rm s}^{-1}}\Bigr)
\Bigl(\frac{30\, {\rm km}}{R}\Bigr)\, .
\label{eq:2}
\end{equation}
Core oscillations can therefore convert accretion power into acoustic power with some efficiency.
In our baseline simulation at the same epoch, $L_{\nu}$(total) is $\sim$${5}\times10^{52}$ erg s$^{-1}$
and $L_{\nu_e/\bar{\nu_e}}$ is $\sim$${2}\times10^{52}$ erg s$^{-1}$.  In contrast,
$L_{\nu_e/\bar{\nu_e}}$$\times\tau_{\nu}$ (where $\tau_{\nu}$ is the gain-region optical depth
to $\nu_e/\bar{\nu_e}$ neutrino absorption) is $<$${10}^{50}$ erg s$^{-1}$.
These power/luminosity comparisons, also depicted in Fig. \ref{lum}, 
highlight the problem, and its potential solution. 

Importantly, what we see in our simulations is the {\it self-excitation} of 
the core g-mode; the inner core plus inner accretion region is a self-excited oscillator.
The interaction between the sound pressure and the accretion streams (see, e.g., Fig. \ref{fig_ent_late})
that results in an anticorrelation at late times between the angular 
positions of the accretion streams and the average direction of the emitted sound, 
and in the emergence of the $\ell=1$ mode, are clear features of our simulations. However, 
the flow fields are quite complicated and variable. What we have yet to 
determine is whether the sound modulates the exciting accretion streams
to create a feedback loop with gain and a time delay, the classic context
for a feedback amplifier, quartz crystal oscillator, acoustic feedback oscillator, klystron, 
or edgetone oscillation.  This possibility requires further exploration, but is intriguing.

The core oscillations are acting like a transducer for the conversion of the gravitational energy
of accretion into acoustic power.  The oscillation continues as long as the accretion
continues, ensuring that acoustic power persists until accretion stops.  This may be one means by
which a supernova, and its energy, are self-regulating, given a progenitor structure.  The halt of accretion
is one manifestation of explosion.  In fact, the accretion streams, sequestered predominantly
on one side of the inner core, pump up the core oscillation and store energy in it like a battery
or capacitor.  Even after accretion subsides the energy in the core motion
will continue to be damped by acoustic radiation, neutrino viscosity, and gravitational
radiation, the former being dominant.  However, as the gas pressure around the
core decreases due to explosion, because of the increasing impedance mismatch
between the core and the expanding mantle, the acoustic damping time
is bound to increase. We have yet to calculate this effect, but the core may oscillate
for a ``long time."  However long it takes, most of the oscillation energy will
eventually be ``discharged" sonically into the supernova explosion.

The angular distribution of the emitted sound is fundamentally aspherical.   
The actual modal periods depend upon the nuclear EOS used,
and will be different when general relativity is incorporated into the calculations.
Relativity will decrease the period and increase the pressures that surround the core, but 
it will also decrease the core size.  Therefore, how relativity will 
alter the acoustic power is not yet clear. Furthermore, how
the accretion streams will behave in 3D, where the pattern of infalling plumes and its evolution
are as yet unknown, remains one of the major uncertainties of our acoustic model.  
A rotation axis may set a direction and render important aspects of a 3D
simulation like those in a 2D simulation.  This is what is suggested in the linear
analysis of Miralles et al. (2004) and what one would conjecture from the
generic alignment effects on convective plumes of the S\o{lberg}-Hoiland instability.
Otherwise, there may be a spectrum of modes that
are not degenerate in azimuth and in ``magnetic quantum number" $m$.
In addition, though we did not see much power in $\ell=2$ modes, they may contribute more
in the real 3D situation.  Nevertheless, from our calculations, 
uniquely allowing as they do core motion on a 2D grid, the excitation of non-radial 
core g-modes seems to be a natural consequence of the later stages of stellar 
collapse.  

\section{The Role of Neutrinos}
\label{neutrinos}

In the baseline simulation we have performed for this paper, despite the fact that
it involves an 11-M$_{\odot}$ progenitor with a steep outer density profile 
and despite the appearance and strengthening of neutrino-driven convection in the 
first $\sim$200 ms after bounce, we do not witness a neutrino-driven explosion.  
This does not mean that the neutrino-driven explosion mechanism is ruled out.  
It may be that different input physics, different progenitors, an improved numerical technique, 
or not employing the MGFLD approximation could result in a 
neutrino-driven explosion.  Since they should accrete their 
outer boundaries quickly within less than $\sim$100 ms after bounce, ONeMg 
white dwarfs that achieve the Chandrasekhar mass and collapse due to 
accretion from a companion (accretion-induced collapse, AIC) can't avoid exploding in the 
early post-bounce stages before the shock instability and core acoustic sources 
turn on.  Nevertheless, with our fully 2D multi-group code, general
Poisson solver, and adequate energy-group and spatial resolution, 
shock stagnation is prolonged until long after the first peak in the average shock radius (near 150-200 km)
and its subsequent initial decay.  In the past, this peaking and decay were  
taken as signatures of a fizzle.

However, if our result is an adequate representation of Nature, then what 
is the role of neutrinos in the more-delayed sound-driven mechanism we have 
discovered?  First, neutrinos still deposit energy in a gain region.  Figure \ref{lum}
depicts not only the neutrino luminosities, but the neutrino
power deposited in the gain region as a function of time.  Hence, neutrinos will certainly
contribute a fraction of the explosion energy.  Second, from the work of Foglizzo
and collaborators and of Blondin, Mezzacappa, \& DeMarino (2003), we expect that
the shock instability requires a large shock height and a large ratio between 
the speeds of sound at the core and at the shock.  Neutrino-driven 
convection results in larger shock radii after the stalling of 
the shock than obtain without neutrino-driven convection. Nevertheless, the shock
instability can't commence until the inner core slows its quasi-static contraction,
a phenomenon slaved to neutrino and lepton losses.  Third, neutrino heating
of the accreted and shocked material that goes through the gain region is 
one reason the entropies around the core increase after $\sim$200 ms 
after bounce.  The other reason is enhanced, multiple shock heating due 
to the shock instability.  Higher entropies and larger entropy perturbations 
result in larger overpressures during the earlier turbulence-driven 
acoustic flux phase seen in our simulations before the core oscillation becomes vigorous.
Fourth, neutrinos are responsible for the failure of the 
direct mechanism for progenitors less massive than $\sim$20 
M$_{\odot}$\footnote{For progenitors more massive, nuclear photodissociation alone 
can stall the shock.}.  If this mechanism succeeded, then the gravitational mass of 
the residual neutron star very likely would be too small to explain
observed pulsar masses.  Fifth, the absorption of $\nu_e$
and $\bar{\nu}_e$ neutrinos by the ejecta may be responsible for ensuring
that its Y$_e$ is not too small to be consistent with nucleosynthesic 
constraints (Frohlich et al. 2005; Pruet et al. 2005).  Furthermore, 
Y$_e$s above 0.5 that are possible only due to an excess of $\nu_e$ 
absorption over $\bar{\nu}_e$ absorption might be needed to explain the
rp-process isotopes.  However, we find that the mass of the core ejecta in
our acoustic-power-driven explosion is less than 0.01 M$_{\odot}$ and
that there may not be as much of a problem with anomalously neutron-rich 
ejecta for the lowest-mass massive stars.  Finally,  neutrino cooling
of the multiple shocks emanating from the core slightly diminishes the acoustic
energy that in our scenario ultimately powers the supernova.  
Determining the correct magnitude of this negative effect will be important and might require 
better neutrino transfer and better weak-interaction physics  
than we have incorporated into our 2D MGFLD approach.

In the calculations we present in this paper, neutrino transport and weak-interaction physics
are crucially important, but are not the critical factors in the explosion
mechanism.  Anisotropic neutrino radiation pressure may in some circumstances
1) affect the core oscillations, and 2) contribute an integrated net impulse
to the PNS and, therefore, a kick.  However, in our baseline model, 
with the single progenitor on which we focus, such effects are subdominant 
and the direct role of neutrinos is subordinate to the other 
factors described in \S\ref{sim} and \S\ref{core}.

\section{Discussion and Conclusions}
\label{conclusion}

In this paper, we present a new mechanism for core-collapse supernova explosions
that focusses on the acoustic power generated in the core region. The strength,
radiation pattern, and character of the emergent sound are influenced by the
shock instability that arises after $\sim$200 ms, but this instability is not
the agent of explosion.  Rather, it is the acoustic power generated first in the 
turbulent region around the inner PNS core and then through the excitation and sonic
damping of core g-mode oscillations.  An $\ell=1$ mode grows at late times to
be prominent around $\sim$500 ms after bounce, though it is in evidence as early
as $\sim$300 ms after bounce. In our Newtonian simulation, its 
period is $\sim$3 ms. At the end of the calculation, this core g-mode 
contains $\sim$$10^{50}$ ergs, is radiating sound into the 
exploding mantle at a rate near $\sim$$10^{51}$ erg s$^{-1}$, and has a 
Q-value of $\sim$200.  At late times, but before explosion, the accreting 
protoneutron star is a self-excited oscillator, ``tuned" to the most easily excited
core g-mode.  After the acoustic power becomes strong, the average direction 
of the radiated sound and the angular positions of the exciting plumes outside 
the core are anti-correlated.  The possibility that the PNS-core/accretion-stream system 
is a feedback amplifier is an intriguing notion suggested by 
our simulations, but that has yet to be properly explored.  Since core $\ell=0$ modes are not
excited to an appreciable degree, the driving acoustic radiation pattern 
is fundamentally anisotropic and the initial phase of the explosion 
is {\it unipolar}.  In addition, due to the extreme breaking of spherical symmetry,
this model manifests simultaneous accretion and explosion.  
The ultimate source of the acoustic power 
is the gravitational energy of infall and the core oscillation
acts like a transducer to convert this accretion energy into sound.

An advantage of the acoustic mechanism is that acoustic power does 
not abate until accretion subsides, so that it is available as long 
as it is needed.  This is not the case for neutrinos, whose luminosities
and mantle heating rates inexorably decrease at late times.   
Hence, this may be the long-sought-after self-regulating mechanism of 
the supernova energy and, being a function of the accretion rate,
is determined mainly by the progenitor density structure,
with some ambiguity due to the sensitivity to initial conditions
in chaotic flows and due to rotation. Furthermore, unlike 
neutrinos, sound pulses that steepen into shock waves
are almost completely absorbed in the baryonic outer mantle.  Moreover,
sound deposits not only energy, but momentum.  Ironically, 
the large accretion rates that were thought to inhibit
explosion in the neutrino-driven mechanism are not a disadvantage in the
acoustic model. In fact, the very accretion that might be inhibiting 
the neutrino mechanism facilitates the acoustic mechanism, but at later
times.  Furthermore, by their nature the accretion funnels constitute a disproportionate
share of the infalling material.  The result is that while at the same time 
these streams are exciting the core oscillations
accretion from the other directions is weaker, thereby presenting
less of an obstacle in those directions to eventual explosion. 
In our calculation with the 11-M$_{\odot}$ progenitor, the delay to
explosion is $\sim$550 ms. This is longer than the delay traditionally
associated with the onset of multi-D neutrino-driven explosions ($\sim$200$-$300 ms).

The baryon mass of our remnant neutron star is $\sim$1.42 M$_{\odot}$, with a 
gravitational mass near $\sim$1.3 M$_{\odot}$. This is close 
to what is canonically expected from measured values. Had the core 
exploded much earlier, the mass remaining might have been uncomfortably lower.


The successive superposition of multiple shocks and the shock-shock interactions
during the shock instability phase lead naturally to higher entropies in a fraction
of the supernova ejecta.  In our baseline calculation, entropies higher than 300
were achieved.  High entropies may be a partial consequence of the thin accreting envelope, 
suggesting that higher-mass progenitors would not so easily yield such high
numbers.  Since the supernova energy, kick velocities, and ejecta entropies
are all dependent in our acoustic model on the accretion regime, these quantities
may in fact be correlated along the progenitor mass continuum, with lower-mass
progenitors having low explosion energies, some r-process nucleosynthesis, and weak kicks,
and higher-mass progenitors having higher explosion energies, no r-process, and 
significant kicks.  Observations do suggest that lower mass massive stars 
are the likely sites of the r-process (Mathews, Bazan, \& Cowan 1992) and that
supernova energies might span a wide range (Hamuy 2003). Be that as it may, 
the progenitor dependence of the supernova explosion systematics is an important aspect of
the supernova story and one that we have yet to explore in the context 
of this new mechanism.   

Though we find that an $\ell=1$ g-mode eventually 
dominates, $\ell=2$ and $\ell=3$ modes and harmonics
are in evidence and there is also likely to be nonlinear mode-mode coupling.  
Unlike $\ell=1$ modes, $\ell=2$ modes will generate 
gravitational radiation and will do so at characteristic frequencies (!)
that is a function of the EOS, relativity, and the PNS structure.  
Using the results from earlier quasi-static PNS cooling calculations, Ferrari et al. (2003)
calculated its g-mode frequencies and their evolution.  Including
as they do general relativity, Ferrari et al. derive frequencies for the 
fundamental g-modes that are slightly higher than we derive using
our Newtonian code, and find that these frequencies, after an initial rise, evolve to lower 
values.  Our calculations recapitulate exactly this initial rise (Fig. \ref{fftp})
and subsequent decay after $\sim$670 ms (not shown in Fig. \ref{fftp}).
The total energy radiated in gravitational waves from this non-rotating 11-M$_{\odot}$
model is $\sim$10$^{-8}$ M$_{\odot}$ equivalent (Ott et al. 2005, in preparation).
Hence, in the excitation of normal modes in the supernova context, 
we may have a direct signature of core physics and supernova phenomenology.

Our calculations are Newtonian and were performed in 2D. 
Relativity and a different nuclear EOS (Ferrari et al. 2003) will change the 
frequencies of the core modes and the efficiency of sound generation,
the latter in ways that are not yet obvious.  
Three-dimensional simulations are likely to be different from 2D simulations,
manifesting more realistic plume structures. The 2D/3D difference is
the major unknown and uncertainty in our scenario.  However, it would be 
difficult to suppress in 3D the excitation of core g-modes and the generation
of acoustic radiation that in 2D becomes so pronounced.  Furthermore, 
even slight rotation may set a natural axis for the evolution in 3D.   
Otherwise, more subtle initial conditions will break the 
symmetry.

\begin{figure}
\plotone{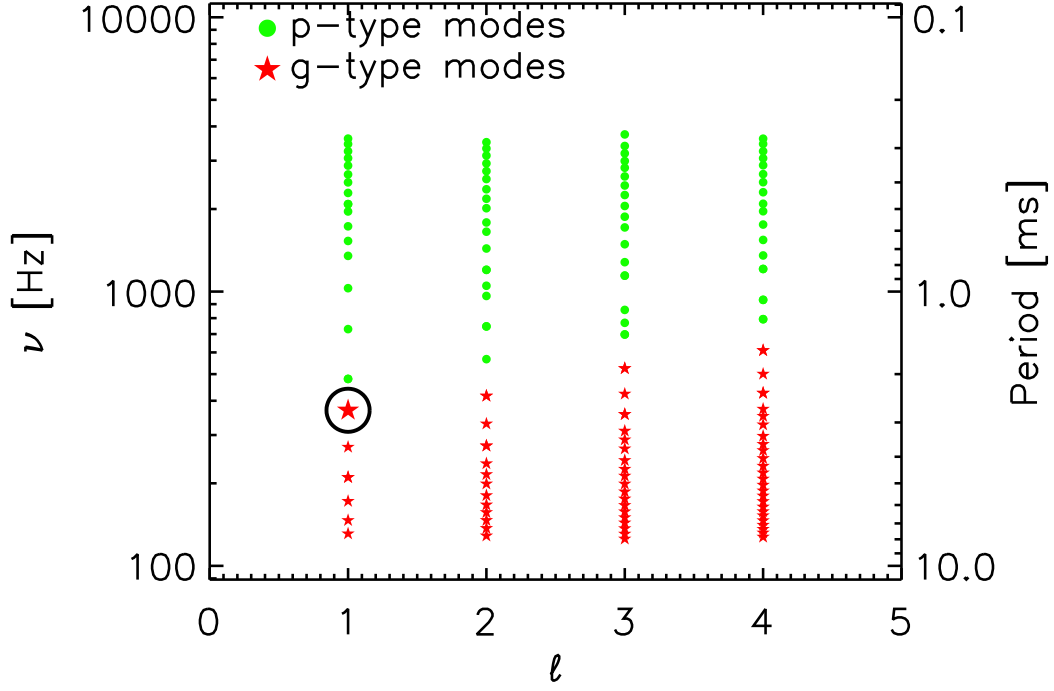}
\caption{Frequency, $\nu$, versus the index, $\ell$, of the spherical-harmonic 
   functions $Y_{{\ell},0}(\theta,\phi)$ for the analytic core g-modes (red stars)
   and p-modes (green circles)
   obtained from a spherical average of the full 2D
   simulation profiles at 500 ms after bounce. The corresponding
   periods ($P=1/\nu$) are given on the right axis.
   Nearly all modes have some g-type nodes and p-type nodes.
   The $\ell =1$ mode highlighted with the circle corresponds
   to the mode with predominantly g-mode character which has
   been most easily excited in our simulation.
\label{modal}}
\end{figure}

\begin{figure*}
\plotone{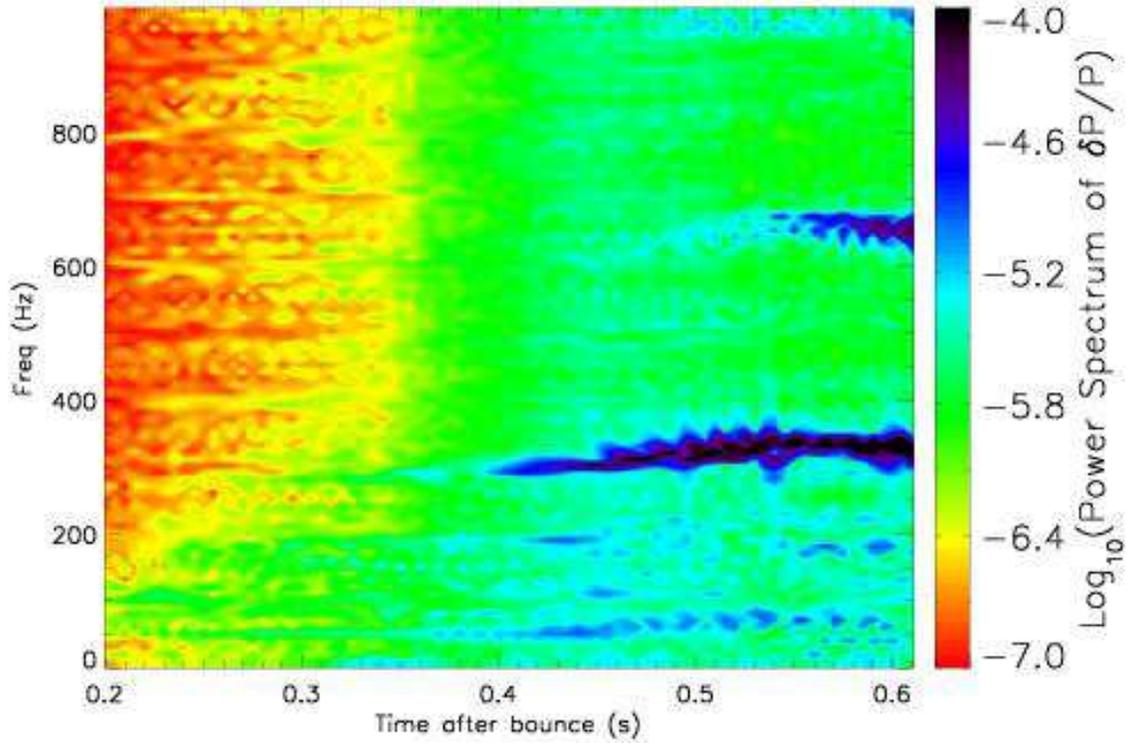}
\caption{
Colorscale of the power spectrum of the fractional pressure variation
($P(R,\theta)-<P(R,\theta)>_\theta) / <P(R,\theta)>_\theta$
at a radius $R = 30$\,km, as a function of time and frequency.
For each time $t$, a power spectrum is calculated from a sample of
time snapshots covering $t\pm$50\,ms, at a resolution of 0.5\,ms.
Note the emergence of power in the $\sim$330 Hz ($\equiv\,$3 ms) g-mode,
as well as the strengthening $\ell=$2 harmonic mode near $\sim$675 Hz 
at late times. The latter is of relevance for gravitational 
radiation emission.
}
\label{fftp}
\end{figure*}

We have also performed a simulation that restricts the hydrodynamics 
to a 90-degree wedge, with a reflecting boundary at the equator,
as opposed to operating in the full 180-degree domain. In this case, we see that 
the $\ell=1$, 25-30 ms wobble due to the shock instability is completely
suppressed.  However, instead of arising near 200-250 ms after bounce, 
the shock instability starts later, around 450-500 ms after bounce, and in 
an $\ell=2$ mode.  Before that, the shock maintains a roughly spherical
shape.  Though the delay is much longer, the shock executes 
small-amplitude, long-period $\ell=2$ motions until roughly $\sim$650 ms
after bounce, at which time the shock instability becomes vigorous.
Just before that, near $\sim$600 ms, the PNS starts to radiate sound waves
from a predominantly $\ell=2$ core g-mode with a $\sim$400 Hz 
frequency (see Fig. \ref{modal}).  Clearly, 90-degree calculations suppress an important 
component of the supernova story, but just as clearly, the same instability 
and oscillation phenomena, though dominated by $\ell=2$ morphology, eventually 
arise.  It just takes longer for $\ell=2$ modes to grow.

It is important to list the reasons the acoustic phenomena we have
identified and presented in this paper were not seen before.  First, most
calculations were stopped after the shock radius first subsided around 200-300 ms after bounce, 
but before the shock instability was much in evidence, and before
turbulence around the core could generate significant acoustic power.
Second, those calculations that were not stopped early were continued because
they experienced an early neutrino-driven, multi-D explosion.  Such an explosion
arose either naturally from the particular code being used, or was artificially produced. 
If the explosion commences early, the PNS core oscillations are not excited 
to useful amplitudes and the shock instability is more mild. Third,
and most importantly, to date all other grid-based supernova codes 
have conducted calculations either with the cores excised, handled
in 1D, or on a $\sim$90$^{\circ}$ wedge, thereby completely
suppressing core oscillations and the resulting $\ell=1$ acoustic flux.  Such procedures can even
muffle the acoustic flux generated in the turbulent inner ``convective" zones. 
As a result of some combination of the reasons above, no previous
supernova simulations, before those using VULCAN/2D, could have 
discovered the acoustic driving and core oscillation mechanism.

Therefore, one key to the discovery of this potentially important mechanism
was the computational liberation of the inner core to execute
its natural multi-dimensional motions.  Another key was patience
to perform the simulations to very late times.
It may be that better neutrino and weak-interaction physics,
full 3D radiation/hydrodynamic simulations, 
multi-group/multi-{\it angle} simulations, a new suite of progenitor
models, the use of other massive-star progenitors (we have here studied only one), 
or some qualitatively important flaw in our approach or implementation
will alter our conclusions here. The role of rotation must be explored,
for both the neutrino and the acoustic models.
Interestingly, when magnetic fields
are incorporated into the calculations, the generation of Alfv\'en waves
at the core oscillation frequency might prove to be another 
power source for the supernova (S. Woosley, private communication).  

The multi-D neutrino-driven 
mechanism may still obtain, and likely does so for AIC systems.
Moreover, it may be that some supernovae explode by the neutrino mechanism\footnote{See the 
papers by Janka et al. 2005ab for the suggestion that the neutrino mechanism might still
obtain for the lowest mass massive stars when the full 180$^{\circ}$ computational domain (even without
liberating the core to execute oscillations) and state-of-the-art Boltzmann transport are incorporated.
These authors seem to see the onset of explosion for an 11.2-M$_{\odot}$ model from Woosley, Heger, \& Weaver (2002).
Interestingly, this model has a small iron core mass of $\sim$1.25 M$_{\odot}$,  
a density cliff near 1600 km, and a much steeper mantle density gradient around 1000-2000 km than
the 11-M$_{\odot}$ model from Woosley \& Weaver (1995) upon which we focus in this paper.},
while others, if the neutrino mechanism fizzles, explode 
by the acoustic mechanism or an MHD-jet mechanism (Akiyama
et al. 2003), though the latter requires very rapid rotation
that may not be available in the generic core-collapse supernova context\footnote{However, such
rapid rotation might enable the MHD scenario for hypernovae and 
long-duration gamma-ray bursts.}.  
However, though unlikely to be correct in detail, our
calculations and analysis may be pointing to important new
phenomena in the theory of core-collapse supernovae that have hitherto
been overlooked.  Much remains to be done, including the determination of
the progenitor-mass and -profile dependence, explorations in 3D with neutrino transport,
and verification and scrutiny by other groups.  Nevertheless, we believe
that the new lines of investigation we have opened up in this paper, and
the potential of the acoustic mechanism interpreted broadly, should stimulate many 
future studies in this central problem in theoretical astrophysics.

\acknowledgments

We thank Rolf Walder, Thierry Foglizzo, Todd Thompson, Itamar Lichtenstadt,
Stan Woosley, Casey Meakin, Chris Fryer, 
and John Blondin for fruitful discussions and their insight. 
We acknowledge support for this work
from the Scientific Discovery through Advanced Computing 
(SciDAC) program of the DOE, grant number DE-FC02-01ER41184,
and from the NSF under grant AST-0504947.
E.L. thanks the Israel Science Foundation
for support under grant \# 805/04, and C.D.O. thanks the Albert-Einstein-Institut
for providing CPU time on their Peyote Linux cluster.
This research used resources of the National 
Energy Research Scientific Computing Center, which is supported by the
Office of Science of the U.S. Department of Energy under Contract No. DE-AC03-76SF00098.
Finally, we thank Don Fisher, Youssif Alnashif, and Moath Jarrah
for their help generating both the color stills and the movies associated with this work.


\clearpage

\clearpage

\clearpage

\clearpage

\clearpage

\clearpage

\clearpage

\begin{thebibliography}{}


\bibitem[Akiyama et al. 2003]{akiyama:03}
Akiyama, S., Wheeler, J.C., Meier, D.L., \& Lichtenstadt, I. 2003,
\apj, 584, 954

 2003ApJ...584..954A

\bibitem[Arnett 1966]{arnett} Arnett, W.D. 1966, Can. J. Phys., 44, 2553

\bibitem[Bethe \& Wilson 1985]{bethe}
Bethe, H. \& Wilson, J.~R.~1985, \apj, 295, 14

\bibitem[Blondin, Mezzacappa, \& DeMarino 2003]{blond03} Blondin, J.M., 
Mezzacappa, A., \& DeMarino, C. 2003, \apj, 584, 971


\bibitem[Bruenn 1985]{BR1} Bruenn, S.W. 1985, \apjs, 58, 771

\bibitem[Bruenn \& Dineva 1996]{dineva} Bruenn, S.W. \& Dineva, T. 1996, \apj,
458, L71

\bibitem[Bruenn et al. 2005]{bruenn2005} Bruenn, S.W., Raley, E.A., \&
Mezzacappa, A. 2005, astro-ph/0404099

\bibitem[Buras et al. 2003]{buras2003} Buras, R., Rampp, M., Janka, H.-Th., \& Kifonidis, K. 2003,
\prl, 90, 241101

\bibitem[Buras et al. 2005]{buras2005} Buras, R., Rampp, M., Janka, H.-Th., \& Kifonidis, K. 2005,
astro-ph/0507135





\bibitem[Burrows,~Hayes,~\&~Fryxell 1995]{bhf}
Burrows, A., Hayes, J., \& Fryxell, B.A.~1995, \apj, 450, 830

\bibitem[Burrows \etal  2000]{B2} Burrows, A., Young, T., Pinto, P., Eastman,
        R. \& Thompson, T. 2000, \apj, 539, 865



\bibitem[Burrows \& Thompson 2004]{bur04f} Burrows, A. \& Thompson, T.A. 2004, ``Neutrino-Matter Interaction 
Rates in Supernovae: The Essential Microphysics of Core Collapse,"
in {\it Core Collapse of Massive Stars},
ed. C. Fryer (Kluwer Academic Press; 2004), p. 133.

\bibitem[Colgate \& White 1966]{colgate} Colgate, S.A. \&  White, R.H. 1966, \apj, 143, 626

\bibitem[Dessart et al. 2005]{dessart} Dessart, L., Burrows, A., Livne, E., \& Ott, C.D. 
2005, submitted to \apj, astro-ph/0510229


\bibitem[Ferrari et al. 2003]{ferrari} Ferrari, V., Miniutti, G., \& Pons, J.A. 2003, \mnras, 342, 629

\bibitem[Foglizzo 2001]{fog01} Foglizzo, T. 2001, \aa, 368, 311

\bibitem[Foglizzo 2002]{fog02} Foglizzo, T. 2001, \aa, 392, 353

\bibitem[Foglizzo \& Tagger 2000]{fogt00} Foglizzo, T. \& Tagger, M. 2000, \aa, 363, 174

\bibitem[Foglizzo, Galletti, \& Ruffert 2005]{fog5a}  Foglizzo, T., Galletti, P., \& Ruffert, M. 2005, \aa, 435, 397

\bibitem[Foglizzo, Scheck, \& Janka 2005]{fog05} Foglizzo, T., Scheck, L., 
\& Janka, H.-T. 2005, astro-ph/0507636

\bibitem[Frohlich et al. 2005]{froh}  Frohlich, C., Hauser, P., Liebend\"{o}rfer, M., Martinez-Pinedo, G.,
Thielemann, F.-K., Bravo, E., Zinner, N.T., Hix, W.R., Langanke, K., Mezzacappa, A., 
\& Nomoto, K. 2005, astro-ph/0410208


\bibitem[Fryer \& Warren 2002]{fryer2002} Fryer, C.L. \& Warren, M. 2002, \apj, 574, L65

\bibitem[Fryer \& Warren 2004]{fryer2004} Fryer, C.L. \& Warren, M. 2004, \apj, 601, 391

\bibitem[Goldreich \& Keeley 1977]{Goldreich1977} Goldreich, P., \& Keeley, D.A. 1977,
\apj, 212, 243

\bibitem[Goldreich \& Kumar 1988]{Goldreich1988} Goldreich, P., \& Kumar, P. 1988,
\apj, 326, 462

\bibitem[Goldreich \& Kumar 1990]{Goldreich1990} Goldreich, P., \& Kumar, P. 1990,
\apj, 363, 694

\bibitem[Hamuy 2003]{hamuy} Hamuy, M. 2003, \apj, 582, 905





\bibitem[Herant et al. 1994]{herant}
Herant, M., Benz, W., Hix, W.R., Fryer, C.L., \& Colgate, S.A. 1994, \apj, 435, 339


\bibitem[Hoffman et al. 1996]{hoff96} Hoffman, R.D., Woosley, S.E., 
Fuller, G.M., \& Meyer, B.S. 1996, \apj, 460, 478




\bibitem[Janka \& M\"uller 1996]{muller96} Janka, H.-T. \& M\"{u}ller, E. 1996, \aa, 306, 167

\bibitem[Janka,~Buras,~\&~Rampp 2003]{janka2003} Janka, H.-T., Buras, R., \& Rampp, M. 2003, Nucl. Phys. A, 718, 269


\bibitem[Janka et al. 2005a]{janka05} Janka, H.-T., Buras, R., Kifonidis, K., Marek, A., 
\& Rampp, M. 2005, in Cosmic Explosions, On the 10th Anniversary of SN1993J. 
Proceedings of IAU Colloquium 192, edited by J.M. Marcaide and 
Kurt W. Weiler, Springer Proceedings in Physics, vol. 99. (Berlin: Springer), p.253 (astro-ph/0401461)

\bibitem[Janka et al. 2005b]{janka05b} Janka H.-Th., Buras R., Kitaura Joyanes F.S., Marek A.,
Rampp M., Scheck L. 2005, ``Neutrino-driven supernovae: An accretion instability in a
nuclear physics controlled environment," in Proceedings of the 8th International Symposium on Nuclei in the
Cosmos, Vancouver, Canada, July 19--23, 2005, Nuclear Physics A, 758, 19--26





\bibitem[Landau \& Lifshitz 1959]{landau} Landau, L.D. \& Lifshitz, 
E.M. 1959, Fluid Mechanics, Pergamon Press Ltd., Oxford



\bibitem[Liebend\"{o}rfer et al. 2001]{lieben2001}
Liebend\"{o}rfer, M., Mezzacappa, A., Thielemann, F.-K., Messer,
O. E. B., Hix, W.~R., \& Bruenn, S.W.~2001, \prd, 63, 103004

\bibitem[Livne (1993)]{livne:93}
Livne, E. 1993, \apj, 412, 634

\bibitem[Livne et al. 2004]{livne04}Livne, E., Burrows, A., Walder, R.,
Thompson, T.A., and Lichtenstadt, I. 2004, \apj, 609, 277


\bibitem[Mathews, Bazan, \& Cowan 1992]{mathews} Mathews, G.J., Bazan, G., \& Cowan, J.J. 1992, \apj, 391, 719

\bibitem[Mayle \& Wilson 1988]{mayle} Mayle, R. \& Wilson, J. R. 1988, \apj,
334, 909




\bibitem[Miralles et al. 2004]{miralles} Miralles, J.A., Pons, J.A., \& Urpin, V. 2004, \aa, 420, 245


\bibitem[Ott et al. 2004]{ott} Ott, C.D., Burrows, A., Livne, E., \& Walder, R. 2004,
\apj, 600, 834

\bibitem[Pruet et al. 2005]{pruet} Pruet, J., Woosley, S.E., Buras, R., 
Janka, H.-T., \& Hoffman, R.D. 2005, \apj, 623, 325  

\bibitem[Rampp \& Janka 2000]{rampp2000} Rampp, M. \& Janka, H.-T. 2000, \apj,
539, L33

\bibitem[Rampp \& Janka (2002)]{rampp20022}
Rampp, M. \& Janka, H.-Th. 2002, \aa, 396, 331


\bibitem[Scheck et al. 2004]{scheck}
Scheck, L., Plewa, T., Janka, H.-Th., Kifonidis, K., \& M\"uller, E. 2004, \prl, 92, 011103

\bibitem[Shen et al. 1998]{shen} Shen, H., Toki, H., Oyamatsu, K., \& Sumiyoshi, K. 1998, 
in "Neutron Stars and Pulsars: Thirty Years after the Discovery," Proceedings of the 
International Conference on Neutron Stars and Pulsars, edited by N. Shibazaki et al.,
(Tokyo, Japan: Universal Academy Press, Frontiers science series, no. 24), p.157

\bibitem[Strack \& Burrows 2005]{strack} Strack, P. \& Burrows, A. 2005, Phys. Rev. D, 71, 
093004, 2005

\bibitem[Swesty \& Myra 2005a]{swesty2005a} Swesty, F.D., \& Myra, E.S. 2005,
astro-ph/0506178

\bibitem[Swesty \& Myra 2005b]{swesty2005b} Swesty, F.D., \& Myra, E.S. 2005,
astro-ph/0507294

\bibitem[Thompson,~Burrows,~\&~Pinto 2003]{Tod1} Thompson, T.A., 
Burrows, A., \& Pinto, P.A., 2003, \apj, 592, 434

\bibitem[Thompson,~Quataert,~\&~Burrows 2005]{tqb} Thompson, T.A., Quataert,
E., \& Burrows, A. 2005,
\apj, 620, 861

\bibitem[Walder et al. 2005]{walder} Walder, R., Burrows, A., Ott, 
C.D., Livne, E., Lichtenstadt, I., \& Jarrah, M. 2005, \apj, 626, 317





\bibitem[Wilson 1985]{wilson85} Wilson, J.R. 1985, in Numerical Astrophysics, 
ed. J. Centrella, J. M. LeBlanc, R. L. Bowers, (Boston: Jones \& Bartlett), p. 422

\bibitem[Wilson \& Mayle 1988]{wm88} Wilson, J.R. \& Mayle, R. 1988, Phys. Rep., 163, 63

\bibitem[Wilson \& Mayle 1993]{wm93} Wilson, J.R. \& Mayle, R. 1993, Phys. Rep., 227, 97

\bibitem[Woosley \& Hoffman 1992]{woos92}
Woosley, S.E. \& Hoffman, R.D. 1992, \apj, 395, 202

\bibitem[Woosley et al. 1994]{woos94} Woosley, S.E., Wilson, J.R., Mathews, G.J., Hoffman, R.D.,
\& Meyer, B.S. 1994, \apj, 433, 229

\bibitem[Woosley \& Weaver 1995]{woosley}
Woosley, S.E. \& Weaver, T.A. 1995, \apjs, 101, 181

\bibitem[Woosley, Heger, \& Weaver 2002]{woosley02}
Woosley, S.E., Heger, A., \& Weaver, T.A. 2002, Rev. Mod. Phys., 74, 1015


\end{thebibliography}
\end{document}